\documentclass[12pt,a4paper]{article}
\pdfoutput=1
\usepackage{jheppub}

\usepackage{amscd,amsmath,amssymb,amsfonts,xspace,mathrsfs,amsthm}
\usepackage{xcolor}

\usepackage{bbold}
\usepackage{latexsym}
\usepackage{graphicx}
\usepackage{tikz-cd}

\numberwithin{equation}{section}

\def\({\left(}
\def\){\right)}
\def\[{\left[}
\def\]{\right]}
\def\<{\left\langle}
\def\>{\right\rangle}

\newcommand{\de}{{\rm d}}
\newcommand{\I}{{\rm i}}

\def\Tr{\,{\rm Tr}\, }
\def\det{\,{\rm det}\, }

\def\sign{{\rm sgn}}

\newcommand{\cA}{\mathcal{A}}
\newcommand{\cC}{\mathcal{C}}

\newcommand{\cF}{\mathcal{F}}

\newcommand{\cQ}{\mathcal{Q}}

\newcommand{\cW}{\mathcal{W}}

\newcommand{\cI}{\mathcal{I}}

\newcommand{\cH}{\mathcal{H}}

\newcommand{\cV}{\mathcal{V}}
\newcommand{\cN}{\mathcal{N}}

\newcommand{\cM}{\mathcal{M}}

\newcommand{\IZ}{\mathbb{Z}}
\newcommand{\IR}{\mathbb{R}}
\newcommand{\IC}{\mathbb{C}}
\newcommand{\IP}{\mathbb{P}}

\def\ba{\bar a}

\newcommand{\be}{\begin{equation}}
\newcommand{\ee}{\end{equation}}
\newcommand{\ben}{\begin{eqnarray}\displaystyle}
\newcommand{\een}{\end{eqnarray}}

\newcommand{\p}{\partial}

\newcommand\bOm{\overline{\Omega}}

\definecolor{varcolor}{rgb}{0.1,0.55,0.25}
\definecolor{functioncolor}{rgb}{0.1,0.35,0.75}
\definecolor{paper_blue}{rgb}{0.3,0.2,0.75}
\definecolor{paper_red}{rgb}{0.65,0.1,0.15}
\definecolor{paper_green}{rgb}{0.05,0.35,0.125}
\definecolor{paper_grey}{gray}{0.375}
\definecolor{perm}{rgb}{0.1,0.45,0.85}
\definecolor{deemph}{rgb}{0.7,0.7,0.7}

\newcommand{\gref}{g_{\rm C}}

\newcommand{\OmS}{\Omega_{\rm S}}

\newcommand{\bOmS}{\bar\Omega_{\rm S}}
\newcommand{\OmT}{\Omega_{\rm T}}
\newcommand{\bOmT}{\bar\Omega_{\rm T}}

\newcommand{\JKres}{\mbox{JK-Res}}

\def\bea{\begin{eqnarray}}
\def\eea{\end{eqnarray}}
\def\be{\begin{equation}}
\def\ee{\end{equation}}
\def\ba{\begin{align}}
\def\ea{\end{align}}
\def\bse{\begin{subequations}}
\def\ese{\end{subequations}}

\def\FC{F_{C,n}}

\newcommand{\mathematica}[4]{\vspace{0.35cm}\noindent\boxed{\begin{minipage}{#2\textwidth}\begin{tabular}{lp{13cm}}{\color{paper_blue}{\scriptsize{\tt In[#1]:}}\raisebox{-0.65pt}{{\scriptsize{\tt=}}}}&{\tt #3}\\{\color{paper_blue}{\scriptsize {\tt Out[#1]:}}\raisebox{-0.65pt}{{\scriptsize{\tt=}}}}&{\tt #4}\end{tabular}\end{minipage}}\vspace{0.35cm}}

\definecolor{varcolor}{rgb}{0.1,0.55,0.25}
\definecolor{functioncolor}{rgb}{0.1,0.35,0.75}
\definecolor{paper_blue}{rgb}{0.3,0.2,0.75}
\definecolor{paper_red}{rgb}{0.65,0.1,0.15}
\definecolor{paper_green}{rgb}{0.05,0.35,0.125}
\definecolor{paper_grey}{gray}{0.375}
\definecolor{perm}{rgb}{0.1,0.45,0.85}
\definecolor{deemph}{rgb}{0.7,0.7,0.7}
\newcommand{\var}[1]{{\tt{\color{varcolor}{\sl#1}}}}

\newcommand{\fun}[1]{{\color{functioncolor}{\tt #1}}}

\title{Quiver indices and Abelianization \\ from Jeffrey-Kirwan residues}

\preprint{arXiv:1907.01354v2}

\author{Guillaume Beaujard,$^{1}$ Swapnamay Mondal,$^{2}$ Boris Pioline$^{1}$
\\
$^1$ {\it Laboratoire de Physique Th\'eorique et Hautes
Energies (LPTHE), UMR 7589 CNRS-Sorbonne Universit\'e,
Campus Pierre et Marie Curie,\\
4 place Jussieu, F-75005 Paris, France} \\

$^2$ {\it International Centre for Theoretical Sciences, \\
Tata Institute of Fundamental Research, Shivakote, \\
Hesaraghatta, Bangalore 560089, India
}

\vspace*{2mm} {\tt e-mail:
\{pioline,beaujard\}@lpthe.jussieu.fr, swapnamay.mondal@icts.res.in
}

\vspace*{-3mm}

}

\abstract{
In quiver quantum mechanics with 4 supercharges, supersymmetric ground states are
known to be in one-to-one correspondence with Dolbeault cohomology classes on
the moduli space of stable quiver representations. Using supersymmetric localization, 
the refined Witten index can be expressed as a residue integral with a specific contour
prescription, originally due to Jeffrey and Kirwan, depending on the stability parameters. 
On the other hand, the physical
picture of quiver quantum mechanics describing interactions of BPS black holes predicts
that the refined Witten index of a non-Abelian quiver can be expressed as a sum 
of indices for Abelian quivers, weighted by `single-centered invariants'. In the case of 
quivers without oriented loops, we show that this decomposition naturally arises from
the residue formula, as a consequence of applying the Cauchy-Bose identity to the 
vector multiplet contributions. For quivers with loops, the same procedure produces 
a natural decomposition of the single-centered invariants, which remains to be elucidated.
In the process, we clarify some under-appreciated aspects of the localization formula.  
Part of the results reported herein have been obtained by implementing the Jeffrey-Kirwan 
residue formula in a public Mathematica code.}

\begin{document}

\maketitle

\section{Introduction}

In four-dimensional quantum field theories or string vacua with 8 supercharges,  BPS states with mutually non-local electromagnetic charges $\gamma_i$ ($ i=1\dots n$), can generically form supersymmetric bound states, or  BPS molecules,
which are stable in some domain in moduli space. 
These bound states can be understood as multi-centered black hole solutions to $\cN=2$ supergravity
in the string theory context \cite{Denef:2000nb,Bates:2003vx}, or as multi-centered Dirac-Julia-Zee dyons in the low energy description of 
$\cN=2$ gauge theories on the Coulomb branch \cite{Lee:1998nv,Denef:1998sv}. In the non-relativistic limit, the interactions between the centers can be described by a supersymmetric quantum mechanics of $n$ particles in $\IR^3$, interacting through electromagnetic, scalar exchange and possibly gravitational 
forces \cite{Bak:1999da,Gauntlett:1999vc,Denef:2002ru,Lee:2011ph,Kim:2011sc}.

Alternatively, in a regime where the BPS constituents can be represented by D-branes, their
dynamics is well described by a supersymmetric gauge theory in 0+1 
dimension \cite{Douglas:1996sw,Fiol:2000pd,Fiol:2000wx,Denef:2002ru,Alim:2011ae,
Alim:2011kw,Cecotti:2012se} with a unitary gauge group $U(N_a)$ for each species
of charge $\alpha_a$ occurring among the $\gamma_i$'s with multiplicity  $N_a$, 
and with $\kappa_{ab}$ chiral multiplets $\phi_{ab,A}$, $A=1,\dots \kappa_{ab}$  in the bifundamental representation $(N_a,\bar{N}_b)$ 
whenever the charges $\alpha_a,\alpha_b$ have non-negative Dirac-Schwinger-Zwanziger product
$\kappa_{ab}=\langle \alpha_a,\alpha_b\rangle$. The field content is graphically represented by a quiver $Q$ with vertices $V_a\in \cV_Q$ associated to the gauge groups $U(N_a)$ and with
arrows $e_{ab,A}\in \cA_Q$ associated to the
chiral multiplets. We shall refer to the
corresponding quantum mechanics with 4 supercharges as quiver quantum mechanics.
The dimension vector $(N_1,\dots, N_K)$ is identified with the total 
electromagnetic charge $\gamma=\sum_{a=1}^K N_a \alpha_a=\sum_{i=1}^n \gamma_i$ 
of the BPS molecule. 

The dependence of the dynamics on the gauge or string theory moduli, which we denote collectively by $z$, is encoded through Fayet-Iliopoulos parameters $\{\zeta_a(z)\}$ associated to each factor 
in the gauge group, as well as through a gauge invariant superpotential $\cW(\phi,z)$ whenever the quiver has oriented closed loops. BPS bound states correspond to supersymmetric
ground states of this quantum mechanics \cite{Denef:2002ru}. In the simplest case where the ranks $N_a$ are coprime and the superpotential is generic, these BPS bound states are in one-to-one correspondence with
Dolbeault cohomology classes on the moduli space $\cM_Q\equiv \cM_Q(\{N_a,\zeta_a\})$ of stable representations,
an algebraic variety of central interest in representation theory (see e.g. \cite{reineke2008moduli}). In particular, the refined Witten index 
$\Omega(\gamma,z,y)=\Tr (-1)^{2J_3} y^{2(I_3+J_3)}$ 
in the supersymmetric quantum mechanics (where $J_3,I_3$ are Cartan generators of
the R-symmetry group $SU(2)_L\times SU_2(R)$)
coincides 
(up to a simple prefactor) with 
the $\chi$-genus of the moduli space $\cM_Q$ \cite{Manschot:2012rx,Lee:2012naa}, 
\be
\label{defchiy}
\Omega(\gamma,z,y) := 
\chi_Q(\{N_a,\zeta_a)\},y) := 
\sum_{p,q=0}^{d}\, (-1)^{p+q-d} y^{2p-d}\, h_{p,q}(\cM_Q)\, 
\ee
which in turn is the specialization at $t=y$ of the Hodge polynomial 
\be
\label{defchiyt}
\Omega(\gamma,z,y,t) := 
\sum_{p,q=0}^{d}\, (-y)^{p+q-d} t^{p-q}\, h_{p,q}(\cM_Q)\, 
\ee
Here $d$ is the complex dimension of $\cM_Q$ and $h_{p,q}$ are the Hodge numbers, such that $\chi_Q$ coincides with $(-1)^d$ times the Euler number $\chi(\cM_Q)$ when $y=1$.
While the Hodge numbers are insensitive to the choice of superpotential away from complex 
codimension-one loci, they depend crucially on the stability parameters $\zeta_a$, and may jump across 
real codimension-one
known as `walls of marginal stability', where $\sum_{a=1}^K n_a \zeta_a=0$ for some positive
integers $n_a$.
The jump is given by a universal wall-crossing formula \cite{ks,Joyce:2008pc,Joyce:2009xv} which has a transparent physical interpretation in terms of the (dis)appearance of multi-centered black hole solutions \cite{Denef:2007vg,Andriyash:2010qv,Manschot:2010qz}.

In general, the direct computation of the Hodge numbers $h_{p,q}(\cM_Q)$ is a difficult task. In the absence
of oriented loops and for primitive dimension vector, $\cM_Q$ is a pure projective variety,
whose Poincar\'e
polynomial, which can be computed by counting points over finite fields \cite{1043.17010}, 
and moreover the cohomology is supported in degree $(p,p)$ \cite{king1995chow}, 
so the $\chi$-genus coincides with the Poincar\'e
polynomial.  Recently, 
building on previous work on two-dimensional gauge linear sigma models \cite{Benini:2013nda,Benini:2013xpa}\footnote{This type of computation was pioneered 
in \cite{Witten:1992xu,Moore:1998et}.}, the  $\chi$-genus 
\eqref{defchiy} for any quiver was computed using the method of supersymmetric localization \cite{Hori:2014tda,Ohta:2014ria,Cordova:2014oxa}, which is closely related to the Atiyah-Bott Lefschetz fixed point theorem.
The result is expressed as a suitable
combination of residues of a rational function $Z_Q(\{u_{a,i}\})$ of the 
(complexified) adjoint scalars $u_{a,i}, i=1\dots N_a$ for each gauge group $U(N_a)$, restricted 
to the Cartan torus. 
For a fixed value of the 
stability parameters $\{\zeta_a\}$, the contributing poles and the order of integration around each of them are determined according to a prescription originally due to Jeffrey and Kirwan \cite{jeffrey1995localization}. Different integration contours arise in different chambers, such that discontinuities across walls of marginal stability are consistent with the wall-crossing formula  \cite{Hori:2014tda}. 
Despite being completely algorithmic, the residue formula becomes quickly unwieldy even for 
moderate ranks, due to a proliferation of possible poles and orders of integration.

Based on the interpretation in terms of a  multi-centered configuration of $n$ BPS black holes with charges
$\gamma_i$, it was suggested in \cite{Manschot:2011xc}, and further elaborated in \cite{Manschot:2012rx,Manschot:2013sya,Manschot:2013dua}, that the Witten index (or $\chi$-genus) of a non-Abelian quiver $Q$ with gauge group $\prod_{a=1}^K U(N_a)$ could be decomposed as a sum of Witten indices $\chi_{Q(\{\gamma_i\})}$ for a family of Abelian quivers $Q(\{\gamma_i\})$ associated to all decompositions of the 
dimension vector $\gamma=\sum_{a=1}^K N_a \alpha_a$ into a sum $\gamma=\sum_{i=1}^m \gamma_i$ where each  $\gamma_i = \sum_{a=1}^K n_{i,a} \alpha_a$ is itself a linear combination
of the basis vectors associated to the nodes of the original quiver, with positive integer coefficients 
$n_{a,i}$. The general formula is known as the  `Coulomb branch formula' from \cite{Manschot:2011xc,Manschot:2012rx}  (see \S\ref{sec_revCoulomb} below for a precise statement) and the coefficients appearing
in front of each $\chi_{Q(\{\gamma_i\})}$ involve a new set of invariants $\bOmS(\gamma_i)$ known as `intrinsic Higgs invariants',  `quiver invariants' or `single-centered invariants' 
\cite{Bena:2012hf,Lee:2012sc,Manschot:2012rx,Lee:2012naa},
which are independent of the stability conditions, and conjecturally depend only on the 
variable $t$ conjugate to $p-q$ in \eqref{defchiyt}, but not on $y$ \cite{Manschot:2012rx}.
These invariants are currently defined in an indirect, recursive way (see \cite{Manschot:2014fua} for a concise explanation of the Coulomb branch formula). This conjecture, if true, gives a powerful way of obtaining the the full Hodge polynomial from the knowledge of $\chi$-genus of $Q$ and of its subquivers.

For quivers without oriented loops and primitive dimension vector however, the
single-centered invariants are known  to have support only on multiples
of the basic vectors $\gamma_i=\ell \alpha_a$, in which case they are simply given by 
$\bOmS(\ell \alpha_a) = (y-1/y)/[\ell (y^\ell-y^{-\ell})]$ for any vertex $V_a\in \cV_Q$. 
The Coulomb branch formula (sometimes known as the 
MPS formula in this restricted setting) then becomes completely explicit, and reduces to 
a sum of the $\chi$-genera for the Abelian quivers $Q(\{\gamma_i\})$, with simple combinatorics
coefficients. This formula was first derived in \cite[App. D]{Manschot:2010qz} based on the 
Reineke formula  \cite{1043.17010} for quivers without loops, and put on a rigorous
mathematical ground in  \cite{ReinekeMPS}.

Our main goal in this paper will be to derive the Abelianization formula for quivers without oriented loops (where the MPS formula is already known to hold) by manipulating the residue formula of \cite{Hori:2014tda}. The key idea is to use the Cauchy-Bose formula
\be
\label{CauchyBose}
\det\frac{1}{\sinh(\mu_i-\nu_j)} =  \frac{\prod_{i<j}\sinh(\mu_i-\mu_j)\, \sinh(\nu_j-\nu_i)}
{\prod_{i,j}\sinh(\mu_i-\nu_j)}
\ee
to decompose the vector multiplet determinant for a given $U(N_a)$ gauge group into a sum over 
conjugacy classes in the permutation group $S_{N_a}$, which  are labelled by 
partitions $\lambda=\sum_{\ell\geq 1} \ell N_{a,\ell}$. Applying \eqref{CauchyBose} to each $U(N_a)$ factor,
leads to a sum over all decompositions of the dimension vector $\gamma=\sum_i \gamma_i$ 
where the $\gamma_i$'s are positive linear combinations of the vectors $\ell \alpha_a$. We shall see that
the contribution $\chi_Q^\lambda$ of permutations\footnote{In fact, all permutations in the same
conjugacy class turn out to contribute equally, naturally leading to the Boltzmann symmetry factor
in the MPS formula.} in the conjugacy class $\lambda$ associated to a decomposition 
$\gamma=\sum_{\ell, a} n_{\ell,a} (\ell \alpha_a)$ reproduce the contribution of the Abelian
quiver $Q(\{\gamma_i\})$ to the MPS formula for  $\chi_{Q}$.
We note that the formula \eqref{CauchyBose} (or rather its rational limit)
was used in a similar context 
in the computation of the index of $\cN=4$ SYM in \cite{Moore:1998et}, and as a tool to construct
grand canonical partition functions for matrix models in  \cite{Moore:1998et,Kazakov:1998ji,Marino:2011eh}. An elliptic version \eqref{CauchyBoseEll} of this equation also allows to implement a similar Abelianization process for computing the elliptic genus two-dimensional gauge linear 
sigma models \cite{Benini:2013xpa}. Finally, we note that 
the  Abelianization formula for quivers was also investigated using toric geometry in \cite{Lee:2013yka}, where a similar but less manipulation of the vector multiplet  determinant
was used.

\medskip

The outline of this work is as follows. In \S\ref{sec_revCoulomb}, we recall basic definitions and facts about quiver moduli spaces, and  briefly review the Coulomb branch formula and the MPS formula.  In \S\ref{sec_residue}, we review the residue prescription of \cite{Hori:2014tda,Benini:2013xpa} for computing the $\chi$-genus $\chi_Q(\{N_a,\zeta_a\})$ of the quiver moduli space $\cM_Q$, recall the relation with the Atiyah-Bott fixed point theorem, examine its value in the attractor chamber, and introduce the Cauchy-Bose formula as a useful way of decomposing the index into a sum over 
partitions of the dimension vector. In \S\ref{sec_ab}, we apply the residue prescription to various
examples of Abelian quivers with or without oriented loop, clarifying the origin of the stability condition for flags contributing to the residue formula. In \S\ref{sec_nonab}, we consider various examples of non-Abelian quivers, and explain the origin of the various terms in the Coulomb branch or MPS formula in terms of conjugacy classes in the Cauchy-Bose formula. In Appendix \ref{sec_math},
we indicate how to reproduce some our results by using the mathematica package  {\tt CoulombHiggs.m} developed by the last-named author, which was originally released along with \cite{Manschot:2013sya},
and has been  extended to include the Jeffrey-Kirwan residue formula and other 
functionalities.

\section{A brief review of the Coulomb branch formula for quivers \label{sec_revCoulomb}}

Let $Q$ be a quiver with vertices $V_a\in \cV_Q$ ($a=1\dots K$), arrows $e_{ab,A}\in \cA_Q$. We assume that there are
no arrows from one vertex to itself, and that all arrows $e_{ab,A}$ between vertices $V_a$ and $V_b$ point in the
same direction. We denote by $\kappa_{ab}$ the number of arrows from $V_a$ to $V_b$, and by 
$-\kappa_{ab}$  the number of arrows from $V_b$ to $V_a$, so that $\kappa_{ab}$ is an antisymmetric matrix with integer entries. Let $\gamma=(N_1,\dots, N_K)$ be a vector of non-negative integers, known as the dimension vector.  The quiver quantum mechanics 
associated to $(Q,\gamma)$ is  a 0+1 dimensional gauge theory with
four supercharges \cite{Denef:2002ru}. It includes vector multiplets for the
gauge group $G=\prod_{a=1}^K U(N_a)$ and $|\alpha_{ab}|$ chiral multiplets transforming
in the bifundamental representation $(N_a,\bar N_b)$ if $\alpha_{ab}>0$, or its complex
conjugate $(\bar N_a,N_b)$ if $\alpha_{ab}<0$. We shall denote the bosonic component
of these chiral multiplets by $\phi_{ab,A,ss'}$, where $1\leq A \leq|\alpha_{ab}|$,
$1\leq s\leq N_a$, $1\leq s'\leq N_b$.  When the quiver has oriented loops, the Lagrangian 
depends on a  superpotential $\cW(\phi)$, which
is a sum of $G$-invariant monomials in the chiral multiplets $\phi_{ab,A,ss'}$.
In addition, the Lagrangian depends
on a real vector $\zeta=(\zeta_1,\dots,\zeta_K)$, known as the stability vector, whose entries are Fayet-Iliopoulos (FI) parameters for the $U(1)$ center in each gauge group $U(N_a)$. Since the diagonal $U(1)$ 
action leaves all fields invariants, we may assume that $\sum_{a=1}^K N_a \zeta_a=0$.
For the purpose of
counting BPS states, the overall scale of the $\zeta_a$'s is also irrelevant, so this vector
can be viewed as a point in real projective space $\IR \IP^K$.

Semi-classically, the quiver quantum mechanics admits two branches of supersymmetric vacua.
On the  Higgs branch,
the gauge symmetry is broken to the $U(1)$ center  by the vevs of the chiral multiplet
scalars $\phi_{ab,A,ss'}$, which are subject to the D and F-term relations
\bea
 \label{Dterm}
&& \sum_{b: \kappa_{ab}>0} \sum_{s'=1\dots b\atop A=1 \dots \kappa_{ab}  } 
\phi_{ab, A, ss'}^*  \, \phi_{ab,A,t s'} 
-  \sum_{b: \kappa_{ab}<0} \sum_{s'=1\dots b\atop A=1 \dots |\kappa_{ab}|} 
\phi_{ba,A, s's}^*  \, \phi_{ba,A,s't} 
 = \zeta_a \, \delta_{st} \quad \forall \, a,s,t  
 \\
 \label{Fterm}
 && {\p W\over \p \phi_{ab,A,ss'}}=0 \quad 
\forall \, a,b,A,s,s'\
\eea
where  $1\leq a\leq K$, $1\leq s,t\leq N_a$ in the first equation while 
$1\leq a,b\leq K$, 
$1\leq s\leq N_a, 1\leq s'\leq N_b, 1\leq
A\leq \kappa_{ab}$ in the second equation, whenever $\kappa_{ab}>0$.
Classical supersymmetric vacua are in one-to-one correspondence with orbits of solutions to 
\eqref{Dterm},\eqref{Fterm}
under the compact gauge group $G=\prod_{a=1}^K U(N_a)$. Equivalently, they are  in one-to-one correspondence with stable orbits of solutions of \eqref{Fterm} under the action of the
complexified gauge group $G_{\mathbb{C}}=\prod_{a=1}^K GL(N_a,\mathbb{C})$, 
where the stability condition is
determined by the vector $\zeta$. The set $\cM_Q$ of supersymmetric vacua thus coincides with the moduli space
of stable quiver representations widely studied in  mathematics (see e.g. \cite{derksen2005quiver,reineke2008moduli}
for entry points in the vast literature on this subject). Quantum mechanically, BPS states on the Higgs branch are harmonic forms on $\cM_Q$, or equivalently Dolbeault cohomology classes \cite{Denef:2002ru}. The group $SO(3)$ associated to
physical rotations in $\IR^3$ acts
on the  cohomology of the Higgs branch via the Lefschetz action generated by
contraction and wedge product with the natural K\"ahler form on $\cM_Q$, induced
from the flat K\"ahler  form on the ambient space $\oplus_{e_{abA}\in \cA_Q} \IC^{N_a} \otimes \IC^{N_b}$.

On the Coulomb branch, the gauge symmetry is broken to the diagonal subgroup
$U(1)^{\sum_{a=1}^K N_a}$ 
and all chiral multiplets as well as off-diagonal vector multiplets are massive. 
After integrating out these degrees of freedom, the diagonal part $\vec r_i$
of the scalars in the vector multiplets must be solutions to 
Denef's equations \cite{Denef:2002ru}
\be
\label{Denefeq}
\forall i=1\dots n\ ,\quad \sum_{j\neq i} \frac{\gamma_{ij}}{|\vec r_i-\vec r_j|}  = c_i\ .
\ee
Here, the index $i$ runs over all $n=\sum_a N_a$ pairs $(a,s)$ with $s=1\dots N_a$, and the corresponding 
$\gamma_{ij}$ and $c_i$
are equal to $\alpha_{ab}$ and $\zeta_a$, in such a way that $\sum_{i=1}^n c_i = \sum_{a=1}^{K}
N_a \zeta_a =0$.
The same equations \eqref{Denefeq} govern the positions of $n$ centers with charges 
$\gamma_i$ such that $\langle \gamma_i,\gamma_j\rangle=\gamma_{ij}$
in $\cN=2$ supergravity \cite{Denef:2000nb,Bates:2003vx}. The space of solutions
modulo common translations  is a phase space   $\cM_n(\{\gamma_{ij}, c_i\})$  of dimension $2n-2$, equipped
with a natural symplectic form (inherited from the symplectic form on the full non-BPS phase space) such that the moment map for spatial rotations is given  by the
total angular momentum $\vec J=\frac12\sum_{i<j}  \gamma_{ij}(\vec r_i-\vec r_j)/|\vec r_i-\vec r_j|$  
\cite{deBoer:2008zn}. BPS states on the Coulomb branch 
are harmonic spinors for the natural Dirac operator on
$\cM_n(\{\gamma_{ij}, c_i\})$ \cite{deBoer:2008zn,Kim:2011sc}.

The Witten index $\Omega(\gamma,z,y)$ defined in \eqref{defchiy} 
counts BPS states on the Higgs branch.
In the
case where the dimension vector $\gamma$ is primitive and the superpotential $\cW$ is generic,
$\cM_Q$ is compact, so $\Omega(\gamma,z,y)$ is a symmetric Laurent polynomial in $y$, which can
be viewed as the character of the Lefschetz action of $SO(3)$ on the cohomology of the quiver 
moduli space $\cM_Q$.
When $\gamma$ is not primitive, $\cM_Q$ is no longer compact, but one can still define the
$\chi$-genus  using intersection cohomology. It is useful to introduce the rational
invariant \cite{Manschot:2011xc}
\be
\label{IntToRat}
\bOm(\gamma,z,y) := 
\bar\chi_Q(\{N_a,\zeta_a)\},y) := \sum_{d|N_a} \frac{y-1/y)}{d(y^d-y^{-d})} \chi_Q(\{N_a/d,\zeta_a)\},y^d) \ ,
\ee
which coincides with $\chi_Q(\{N_a,\zeta_a)\},y)$ when the dimension vector $\gamma=(N_1,\dots, N_K)$ is primitive, but is in general a rational function of $y$ whenever $\gamma$ is not primitive.
The advantage is that $\bar\chi_Q(\{N_a,\zeta_a)\},y) $ satisfies a much simpler wall-crossing
formula than $\chi_Q(\{N_a,\zeta_a)\},y)$ \cite{Joyce:2008pc,Joyce:2009xv,Manschot:2010qz}.

\subsection{The Coulomb branch formula}

The Coulomb branch formula conjecturally expresses the rational index $\bOm(\gamma,z,y)$
in terms of single-centered indices
$\bOm_S(\gamma_i)$ as follows \cite{Manschot:2011xc,Manschot:2013sya,Manschot:2014fua}:
\be
\label{CoulombForm1}
\bar\chi_Q(\{N_a,\zeta_a)\},y)  =  \sum_{\gamma=\sum_{i=1}^n\gamma_i}
\frac{\gref(\{\gamma_i,c_i\},y)}{|{\rm Aut}\{\gamma_i\}|}
\prod_{i=1}^n \bOmT(\gamma_i,y)
\ee
where $\bOmT(\gamma_i,y)$ is constructed in terms of $\OmT(\gamma_i,y)$ by a relation
similar to \eqref{IntToRat}. The `total' invariant  $\OmT(\gamma_i,y)$ is in turn determined in terms of the single-centered
invariants $\Omega_S(\gamma_i,y)$ via
\be
\label{CoulombForm2}
\OmT(\gamma,y)=\Omega_S(\gamma,y)+
\sum_{\gamma=\sum_{i=1}^m m_i \beta_i} H(\{\beta_i,m_i\},y)\prod_{i=1}^m \Omega_S(\beta_i,y^{m_i}).
\ee
In  \eqref{CoulombForm1}, the sum runs  over unordered decompositions of $\gamma$
into a sum of  vectors $\gamma_i=\sum_{a}n_{i,a} \alpha_a$ which are linear
combinations of the basis vectors $\alpha_a$ with positive integer coefficients.
Similarly, in  \eqref{CoulombForm2} the sums run over unordered decompositions of $\gamma$
into sums of  vectors $m_i \beta_i$ with $m_i\geq 1$ and $\beta_i$ a linear
combination of the $\alpha_a$'s with positive integer coefficients\footnote{If one of the constituents
$\beta_i$ is not primitive, all choices $(d m_i,\beta_i/d)$ are counted as distinct contributions.}.
The functions $H(\{\beta_i,m_i\},y)$ are determined recursively
by the so called ``minimal modification hypothesis" (see  \cite{Manschot:2013sya,Manschot:2014fua} for details)
and their role is to ensure that the full refined index $\bOm(\gamma,z)$ is a symmetric Laurent polynomial in $y$.
The function $\gref(\{\gamma_i,c_i\},y)$, known as the Coulomb index,
is the only quantity on the r.h.s. of \eqref{CoulombForm1} which depends on the stability parameters 
$\zeta_a$.
It is defined as the equivariant index of the Dirac operator on the phase space $\cM_n(\{\gamma_i,c_i\})$,
computed by localization with respect to rotations around a fixed axis \cite{Manschot:2010qz,Manschot:2011xc,Manschot:2013sya}. The fixed points
of the action of $J_3$ on $\cM_n(\{\gamma_i,c_i\})$ are collinear black hole solutions,
which are classified by  permutations $\sigma$ of $\{1,2,\dots n\}$,
\be
\label{gCFy}
\gref(\{\gamma_i,c_i\},y)
= \frac{(-1)^{n-1+\sum_{i<j} \gamma_{ij}}}{(y-y^{-1})^{n-1}}
\sum_{\sigma\in S_n}
\FC(\{\gamma_{\sigma(i)},c_{\sigma(i)}\})\,
y^{\sum_{i<j} \gamma_{\sigma(i)\sigma(j)}},
\ee
where the `partial Coulomb index' $\FC(\{\gamma_i\,c_i\}) \in\IZ$ counts (with sign) collinear solutions
for a fixed ordering $x_1<x_2\dots <x_n$ along the axis. A recursive procedure for computing it
was given in  \cite{Manschot:2013sya} and has been implemented in a Mathematica package 
(see \S\ref{sec_math}).

When the Abelian quiver $Q(\{ \gamma_i \})$ constructed from the adjacency matrix
$\gamma_{ij} = \langle \gamma_i,\gamma_j \rangle = \sum_{a,b} n_{i,a} n_{j,b} \kappa_{ab}$
has no oriented loop, the Coulomb index \eqref{gCFy}
coincides with the $\chi$-index $\chi_{Q(\{ \gamma_i \})}$ with stability parameters $c_i=
\sum_a n_{i,a} \zeta_a$, in particular it is a symmetric Laurent polynomial in $y$. 
When $Q(\{ \gamma_i \}$ has oriented closed loops, this relation is lost, and 
the Coulomb index \eqref{gCFy} is in general a rational function. The 
 functions
$H(\{\beta_i,m_i\},y)$ are then adjusted in such a way that the full index $\bOm(\gamma,z)$ obtained
via \eqref{CoulombForm1} is a symmetric Laurent polynomial in $y$, provided  the single-centered
indices $\Omega_S(\gamma_i,y)$ are. The minimal modification hypothesis of \cite{Manschot:2013sya}
gives a unique prescription for computing $H$,
based on the assumption that the missing contributions from the boundary of $\cM_n$ carry the minimal
possible angular momentum. Note that this prescription does not take into account the condition
of absence of closed timelike curves, which is  irrelevant in the context of quiver
quantum mechanics, but needs to be checked by hand for  general supergravity bound states 
(see e.g. \cite[\S 3.2]{Manschot:2011xc} for an example where this condition makes an important difference).

In the special case where the dimension vector 
$\gamma$ is primitive and such that all charge vectors $\gamma_i$ appearing in each decomposition
$\gamma=\sum \gamma_i$ are distinct and primitive (which in particular applies when the original 
quiver $Q$ is Abelian),  the Coulomb branch formula
\eqref{CoulombForm1} simplifies to
\be
\label{CoulombForm3a}
\begin{split}
\Omega(\gamma,z,y) = \sum_{\gamma=\sum_{i=1}^n\gamma_i}\!\!\!
\gref(\{\gamma_i,c_i\},y)\,
\prod_{i=1}^n
\left\{
\Omega_S(\gamma_i,y)
+\!\!\!\!\sum_{\sum_{j=1}^{m_i} \beta_j=\gamma_i}  \!\!\!\!
H_{m_i}(\{\beta_j\},y)\, \prod_{j=1}^{m_i} \OmS(\beta_j,y)  \right\}.
\end{split}
\ee
In this case, the rational functions $H(\{\beta_j\},y)$ are fixed by demanding that the coefficient
of the monomial $\prod_{j=1}^m \OmS(\beta_j,y)$ in $\Omega(\gamma,z,y)$ be a Laurent polynomial in $y$.
Requiring that $H(\{\beta_j\},y)$ are invariant under $y\to 1/y$ and vanish at $y=\infty$ fixes
them uniquely \cite{Manschot:2011xc}. In \cite{Manschot:2012rx} it is conjectured that
the Hodge polynomial \eqref{defchiyt} of $\cM_Q$ can be obtained from the $\chi$-genus
given by \eqref{CoulombForm1} or \eqref{CoulombForm3a} by replacing the argument $y$ in 
$\Omega_S(\gamma_i,y)$ by the fugacity $t$ conjugate to $p-q$.

\subsection{The MPS formula}
\begin{figure}[h]
\begin{center}
\begin{tikzpicture}[scale=2, minimum size=4mm]
\begin{scope}[shift={(-2,0)}]
  \node (b) at ( -1,0) [circle,draw] {$V_b$};
  \node (a) at ( 0,0) [circle,draw,label=below:$N_a$] {$V_a$};
  \node (c)  at ( 1,0) [circle,draw] {$V_c$};
  \draw [->] (b) to node[auto] {$\kappa_{ba}$} (a);
  \draw [<-] (c) to node[auto,swap] {$\kappa_{ac}$} (a);
\end{scope}
\begin{scope}[shift={(2,0)}]
  \node (b) at ( -2,0) [circle,draw] {$V_b$};
    \node (c)  at ( 2,0) [circle,draw] {$V_c$};
  \node (a1) at ( 0,3) [circle,draw] {$W_{a,1,1}$};
  \node (a2) at ( 0,2) [circle,draw] {$W_{a,1,n_1}$};
  \node (a3) at ( 0,.7) [circle,draw] {$W_{a,2,1}$};
  \node (a4) at ( 0,-.3) [circle,draw] {$W_{a,2,n_2}$};
  \node(a5) at ( 0,-2) [circle,draw] {$W_{a,\ell,1}$};
  \node (a6) at ( 0,-3) [circle,draw] {$W_{a,\ell,n_\ell}$};
  \draw [->] (b) to node[auto] {$\kappa_{ba}$} (a1);
  \draw [->] (b) to node[auto] {$\kappa_{ba}$} (a2);
   \draw [->] (b) to node[auto,swap] {$2\kappa_{ba}$} (a3);
  \draw [->] (b) to node[auto,swap] {$2\kappa_{ba}$} (a4);
    \draw [->] (b) to node[auto] {$\ell\kappa_{ba}$} (a5);
  \draw [->] (b) to node[auto,swap] {$\ell\kappa_{ba}$} (a6);
  \draw [<-] (c) to node[auto,swap] {$\kappa_{ac}$} (a1);
  \draw [<-] (c) to node[auto] {$\kappa_{ac}$} (a2);
   \draw [<-] (c) to node[auto,swap] {$2\kappa_{ac}$} (a3);
  \draw [<-] (c) to node[auto,swap] {$2\kappa_{ac}$} (a4);
     \draw [<-] (c) to node[auto,swap] {$\ell\kappa_{ac}$} (a5);
  \draw [<-] (c) to node[auto] {$\ell\kappa_{ac}$} (a6);
  \draw [dotted](a1) to (a2);
      \draw [dotted](a3) to (a4);
          \draw [dotted](a5) to (a6);
    \draw [dotted](0,-.8) to (0,-1.6);
\end{scope}
\end{tikzpicture}
\end{center}
\caption{Blowing up a non-Abelian node of rank $N_a=\sum_\ell \ell n_{a,\ell}$ (on the left) into
$\sum_\ell n_{a,\ell}$ Abelian nodes (on the right). The number of arrows from $W_{a,\ell,m}$
to $V_b$ is $\ell$ times the number of arrows from $V_a$ to $V_b$, where $\ell$ is the level
of the node $W_{a,\ell,m}$. This operation may be performed on any number of non-Abelian nodes.
\label{fig:MPS}}
\end{figure}
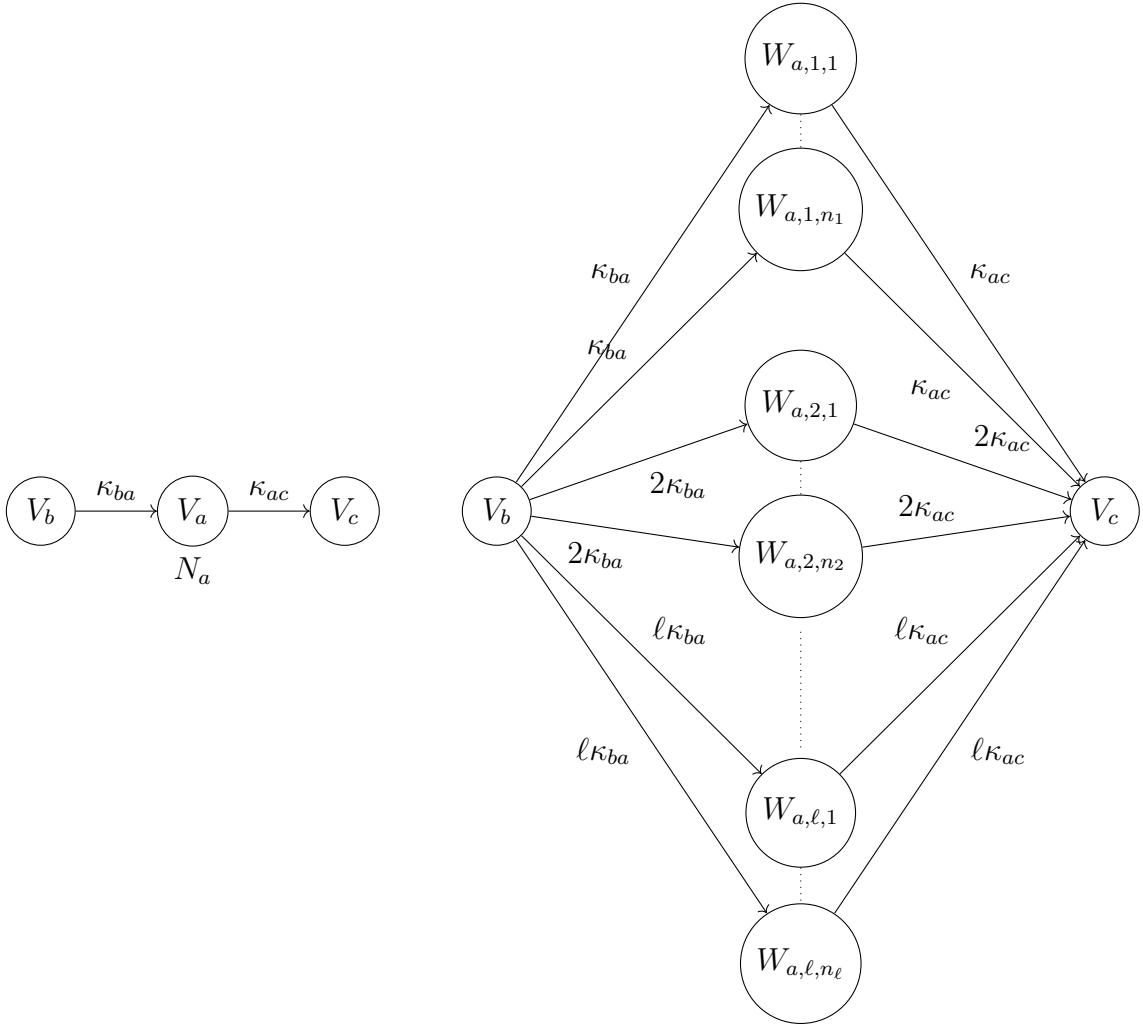

For quiver without oriented loops, the single-centered invariants $\bOm_S(\gamma_i)$
vanish unless $\gamma_i$ is a multiple of a basis vector $\alpha_a$, in which case 
$\bOmS(\ell \alpha_a) = (y-1/y)/[\ell (y^\ell-y^{-\ell})]$. The  Coulomb branch formula 
therefore reduces to
\be
\label{MPS}
\bar\chi_Q(\{N_a,\zeta_a\},y) = \sum_{N_a=\sum\ell \ell n_{\ell a} \atop a=1\dots K}
\chi_{Q(\{n_{\ell a}\})}(\{c_i\},y)
\, \prod_{a=1}^K\, \prod_{\ell} \frac{1}{n_{\ell,a}!}  \left[ \frac{y-1/y}{\ell (y^\ell-y^{-\ell})} \right]^{n_{\ell,a}}
\ee
The factor $\chi_{Q(\{n_{i\ell,a})}(\{c_i\},y)$ is the $\chi$-genus of 
 an Abelian quiver with $n=\sum_{\ell,a} n_{\ell,a}$ 
vertices $\{W_{a,\ell,m}, 1\leq a\leq K, 1\leq m\leq   n_{\ell,a}\}$, which we denote by 
$Q(\{n_{\ell a}\})$  (see Figure \ref{fig:MPS}).
Denoting the nodes of the `blown up quiver'  by $\{W_i, 1\leq i\leq n\}$
using an arbitrary bijection $i \mapsto (a(i),\ell(i),m(i))$, the adjacency matrix of $Q(\{n_{\ell a}\})$ 
is given by $\gamma_{ij}=\ell(i) \ell(j) \kappa_{a(i)a(j)}$, the dimension vector  has entries $N_i=1$  while the stability vector is 
$c_i=\ell(i) \zeta_{a(i)}$, in such a way that $\sum_{i=1}^n c_i = \sum_{\ell,a} \ell n_{\ell,a} \zeta_a=0$.
The positive integer $\ell(i)$ is sometimes called the {\it level}  of the  vertex $W_i$. Physically,
the quiver $Q(\{n_{\ell a}\})$ describes the interactions of $n=\sum_\ell n_{\ell a}$ black holes
with charges  $\gamma_i=\ell(i) \alpha_{a(i)}$, treated as distinguishable particles, so we 
use the notation  the equivalent notation $Q(\{\gamma_i\})$ for the quiver $Q(\{n_{\ell a}\})$.

\medskip

The formula \eqref{MPS} is a special case $S=\{1,\dots, K\}$ of a more general  relation operating only  on a subset $S\subset \{1,\dots, K\}$ of the  nodes:
\be
\label{MPSpartial}
\bar\chi_Q(\{N_a,\zeta_a\},y) = \sum_{N_a=\sum_\ell \ell \,n_{\ell,a} \atop  a\in S}
\chi_{Q(\{n_{\ell,a})\}} (\{c_i\},y)
\, \prod_{a\in S} \prod_{\ell} \frac{1}{n_{\ell,a}!}  \left[ \frac{y-1/y}{\ell (y^\ell-y^{-\ell})} \right]^{n_{\ell,a}}
\ee
where the quiver $Q(\{n_{\ell,a}\})$ is obtained from $Q$ by replacing every vertex $V_a\in \cV_Q\cap S$
with  vertices $W_{a,\ell,m}$, $1\leq m\leq n_{\ell,a}$, carrying level $\ell$ while the vertices $V_a \in V_Q\backslash S$ are unaffected. Again, we denote by $\{W_i\}$ the set of all vertices, and by 
$a(i),\ell(i),N(i)$ their type, level and dimension, equal to $(a,\ell,1)$ for  $W_{a,\ell,m}$ and to $(a,1,N_a)$ for $V_a \in V_Q\backslash S$.
The  adjacency matrix is given as before by $\gamma_{ij}=\ell(i) \ell(j) \kappa_{a(i)a(j)}$,  the stability vector is  $c_i=\ell(i) \zeta_{a(i)}$ and the dimension vector has entries $N_i=N(i)$, in such a way that 
$\sum_{i=1}^n N_i c_i =0$. Clearly, this relation is only non-trivial if the entries in $S$ correspond to non-Abelian nodes, i.e. $N_a>1$ for $a\in S$. Moreover, the relation \eqref{MPSpartial} follows
iteratively from the special case where $S$ contains a single element. In the case where the
dimension vector is primitive, this relation was established in  \cite[App. D]{Manschot:2010qz} based on Reineke's formula \cite{1043.17010} for quivers without oriented loops and with primitive dimension vector. It is worth noting that the proof also works for quivers with oriented loops but with
vanishing superpotential, since  Reineke's formula also holds in that case \cite{Manschot:2012rx}.

\section{Indices from residues \label{sec_residue}}

Using supersymmetric localization techniques, the authors of \cite{Hori:2014tda,Cordova:2014oxa} have found a general prescription for computing the Witten index in any gauged quantum mechanics with two supercharges. In the case of quiver quantum mechanics with four supercharges, 
the  prescription reads:
\be
\label{Omloc}
\bar\chi_Q(\{N_a,\zeta_a\},y) 
= \frac{1}{\prod_{a=1}^K N_a!} \sum_{p} \JKres_p \left( Z_Q(\{u_{a,s}\},y) , \zeta \right) \, 
\ee
Here, the variables $u_{a,s}$ parametrize the complexified Cartan torus $\prod_{a=1}^K (\IC^\times)^{N_a}$ of the gauge group 
$[\prod_a U(N_a)]/U(1)$, the factor $Z_Q(\{u_{a,s}\},y)$ is a 
holomorphic top form on this space and
the sum runs over all poles of $Z_Q(\{u_{a,s}\},y)$. The symbol $\JKres$ denotes a specific prescription, 
originally 
due to Jeffrey and Kirwan \cite{jeffrey1995localization}\footnote{The specific prescription used in
\cite{Benini:2013xpa,Hori:2014tda} and in this paper  can be found in \cite{SzenesVergne}.}, for extracting the residue  
at the corresponding pole. The prescription crucially depends on the stability vector
$\zeta$, and ensures that the result is consistent with the wall-crossing formula. The function
$Z_Q(\{u_{a,s}\},y)$ originates from the one-loop fluctuation determinant of all fields in the 
vector and chiral multiplets,  and is given by 
\be
\begin{split}
\label{Zuhbar}
Z_Q(\{u_{a,s}\},y)  = & \left[\frac{\pi\hbar}{\sin(\pi \hbar)}\right]^{\sum_{a=1}^K N_a-1}
\prod_{a=1\dots K \atop{ s,s'=1\dots N_a \atop s\neq s'}} 
\frac{\sin[ \pi \hbar (u_{a,s'}-u_{a,s}) ]}{\sin[ \pi \hbar(u_{a,s}-u_{a,s'}-1)}\, \\
& \prod_{a,j=1\dots K \atop \kappa_{ab}>0}
\prod_{s=1\dots N_a \atop s'=1\dots N_b}
\prod_{A=1\dots \kappa_{ab}}
 \frac{\sin[ \pi \hbar(u_{b,s'}-u_{a,s} +1-\tfrac{R_{ab}}{2}- \theta_{ab,A})]}
 {\sin[ \pi \hbar(u_{a,s}-u_{b,s'}+\tfrac{R_{ab}}{2}+ \theta_{ab,A})]}\
\, \widetilde{\prod_{a=1\dots K \atop  s=1\dots N_a}} \de u_{a,s}
\end{split} 
\ee
where $y=e^{\i \pi \hbar}$, and $ \widetilde{\prod}$ denotes the omission 
of any one of the $\de u_{a,s}$'s
in the measure (due the decoupling of the diagonal $U(1)$ action). The quantity $R_{ab}$ denotes
the $R$-charge of the chiral fields $\phi_{ab,A,ss'}$, while the parameters $\theta_{ab,A}$
are chemical potentials\footnote{Since the R-charge is defined only up to the addition of flavor symmetries,
one may decide to absorb $\theta_{ab,A}$ into $R_{ab,A}=R_{ab}+2\theta_{ab,A}$. However we find it convenient to separate $R_{ab,A}$ into a flavor invariant part 
$R_{ab}$ and chemical potentials $\theta_{ab,A}$ which may break part of the flavor symmetry. }
 for the $U(\kappa_{ab})$ flavor symmetry permuting
the chiral fields $\phi_{ab,A,ss'}$ with $A=1\dots \kappa_{ab}$.  In the absence of oriented loops, the
assignment of $R$-charges is irrelevant, but if oriented loops are present, it constrains the possible
superpotential, which must be a gauge invariant polynomial in the $\phi_{ab,A,ss'}$ with R-charge 
equal to 2, invariant under flavor symmetries with non-zero potential $\theta_{ab,A}$. In particular,
in order to allow for a generic superpotential, one should set all flavor potentials to zero and 
ensure that the total R-charge for any oriented loop is  equal to 2. 
Clearly, \eqref{Zuhbar} is  invariant under $\hbar\to -\hbar$, or equivalently $y\to 1/y$.

In the limit $ y\to 1$ (or $\hbar\to 0$), and for vanishing chemical potentials $\theta_{ab,A}=0$, the $\chi$-genus reduces to the Euler number, 
given by the sum of residues of 
\be
\begin{split}
\label{Zuhbar1}
Z_Q(\{u_{a,s}\})  = & 
\prod_{a=1\dots K \atop{ s,s'=1\dots N_a \atop s\neq s'}} 
\frac{u_{a,s'}-u_{a,s}}{ u_{a,s}-u_{a,s'}-1}
\prod_{a,b=1\dots K \atop \kappa_{ab}>0}
\prod_{s=1\dots N_a \atop s'=1\dots N_b}
\left[ \frac{u_{b,s'}-u_{a,s} +1-\tfrac{R_{ab}}{2}}{u_{a,s}-u_{b,s'}+\tfrac{R_{ab}}{2}}
\right]^{\kappa_{ab}}
\widetilde{\prod_{a=1\dots K \atop  s=1\dots N_a}} \de u_{a,s} 
\end{split} 
\ee
When $\theta_{ab,A}\neq 0$,  the limit $\hbar\to 0$ instead produces the equivariant Euler number for the action of the Cartan torus of the flavor symmetry.
It is worth noting that \eqref{Zuhbar} arises as the limit $\tau\to\I\infty$ from the elliptic genus of
a two-dimensional gauged linear sigma model with the same matter content, given by a sum
of residues of 
\be
\begin{split}
\label{Zelliptic}
Z_Q(\{u_{a,s}\},y,\tau) = & \left[\frac{2\pi \eta^3\hbar}{\theta_1(\hbar)}\right]^{\sum_{a=1}^K N_a-1}
\prod_{a=1\dots K \atop{ s,s'=1\dots N_a \atop s\neq s'}} 
\frac{\vartheta_1[ \pi \hbar (u_{a,s'}-u_{a,s}) ]}{\vartheta_1[ \pi \hbar(u_{a,s}-u_{a,s'}-1)}\, \\
& \prod_{a,j=1\dots K \atop \kappa_{ab}>0}
\prod_{s=1\dots N_a \atop s'=1\dots N_b}
\left[ \frac{\vartheta_1[ \pi \hbar(u_{b,s'}-u_{a,s} +1-\tfrac{R_{ab}}{2})]}{\vartheta_1[ \pi \hbar(u_{a,s}-u_{b,s'}+\tfrac{R_{ab}}{2})]}\
\right]^{\kappa_{ab}}\,\widetilde{\prod_{a=1\dots K \atop  s=1\dots N_a}} \de u_{a,s} 
\end{split} 
\ee
where $\theta_1(v,\tau)=\theta_1(v)=2 q^{1/8} \sin(\pi v)\, \prod_{n=1}^{\infty}
(1-q^n)(1-q^n e^{2\pi\I v})(1-q^n e^{-2\pi\I v})$. Note however that in 0+1 dimensions, there is no need to  cancel the R-symmetry anomaly, unlike for the elliptic genus of two-dimensional sigma models.

In order to compute the residue in \eqref{Omloc}, it is more efficient to change variables to $v_{a,s}=e^{2\pi\I \hbar u_{a,s}}$, such that \eqref{Zuhbar} may be replaced by 
\be
\label{Zv}
\begin{split}
Z_Q(\{v_{a,s}\},y)  = &
\prod_{a=1\dots K \atop{ s,s'=1\dots N_a \atop s\neq s'}} 
\frac{v_{a,s'}-v_{a,s}}{v_{i,s}/y-y\, v_{a,s'}} 
\\& 
\prod_{a,b=1\dots K \atop \kappa_{ab}>0}
\prod_{s=1\dots N_a \atop s'=1\dots N_b}
\prod_{A=1\dots \kappa_{ab}}
\frac{y^{1-R_{ab}} \,\nu_{ab,A} \, v_{b,s'} - v_{a,s}/y}{v_{a,s}-y^{-R_{ab}}\nu_{ab,A} \, v_{b,s'}}
\widetilde{\prod_{a=1\dots K \atop  s=1\dots N_a}}  \frac{\de v_{a,s}}{v_{a,s}(y-1/y)}
\end{split} 
\ee
where $\nu_{ab,A}=e^{-\I\pi\theta_{ab,A}}$ are flavor fugacities.
In this representation, the symmetry under $y\to 1/y$ requires inverting $v_{a,s}\to 1/v_{a,s}$,
$\alpha_{ab,A}\to 1/\alpha_{ab,A}$. 

\subsection{Review of the JK residue prescription}

 To explain the prescription for identifying and extracting the relevant residues, let us introduce the weight lattice $\Lambda=\IZ^{\sum N_a}$, with basis $e_{a,s}$, $a=1\dots K, s=1\dots N_a$.
To each root of $U(N_a)$ we associate the vector $e_{a,s}-e_{a,s'}$ dual to the hyperplane $H_{a,s,s'}=u_{a,s}-u_{a,s'}-1$. To each bifundamental chiral field 
$\phi_{ab,A,ss'}$
with $\kappa_{ab}>0$ we associate the vector $e_{a,s}-e_{b,s'}$ dual to the hyperplane 
$H_{ab,A,ss'}=u_{a,s}-u_{b,s'}+\tfrac{R_{ab}}{2}+\theta_{ab,A}$. If the flavor chemical potentials $\theta_{ab,A}$ vanish for all $A$, we may
as well consider a single hyperplane $H_{ab,ss'}$ with multiplicity $|\kappa_{ab}|$.
Finally, 
since all fields are neutral under the diagonal $U(1)$ action, we set one Cartan parameter to zero,
e.g.  $u_{1,1}=0$, and disregard the corresponding entry  of all charge vectors. We denote by 
$r=\sum_{a=1}^K N_a-1$ the rank of the reduced lattice $\Lambda/\IZ$.

Now, for each isolated intersection $p$ of hyperplanes, let $\cH_p$ be the list of hyperplanes intersecting at $p$; to construct this list, consider all $r$-plets of hyperplanes, determine their intersection\footnote{For quiver theories, each non-degenerate $r$-plet of hyperplanes  appears to have only one intersection point on the Cartan torus, but this is not true for more general matter content}, collect all non-degenerate intersection points, and  finally for each point in this list, collect all hyperplanes which meet at that point. Let $\cQ_p$ be the corresponding list of charge vectors. We further assume that 
the intersection is projective, i.e. that all vectors in $\cQ_p$ lies in a positive half-space of $\IR^r$. If that is not the case, it is usually possible to perturb the charges $R_{ab}$ so as to resolve a non-projective intersection into multiple projective intersections.

Let $\mathcal{F}_p$ be the list of flags $F=(F_1,\dots,F_r)$ made out of vectors in $\cQ_p$; equivalently, the list of ordered 
$r$-plets $Q_F=(Q_1,\dots Q_r)$ of vectors in $\cQ_p$, subject to the equivalence relation 
$Q_F\sim Q'_F$ if $Q_F Q_F^{'-1}$ is a lower triangular matrix, if the rows of $Q_F$ denote the charge vectors. The space spanned by $Q_j$ with $1\leq i\leq j$ defines the $i$-th graded space $F_i$ in the 
flag $F$. Let $\kappa_{F,i}$ be the sum of all charge vectors\footnote{Here it matters whether we treat
hyperplanes with $|\kappa_{ab}|>1$ as multiple copies with unit multiplicity, or a single copy with multiplicity $|\kappa_{ab}|$. We find in examples that the two prescriptions lead to the same result.}
 in $\cQ_p$ which belongs to $F_i$,
for all $i=1,\dots r$.  Note that $\kappa_{F,i}$  includes contributions from the vectors $Q_1,\dots, Q_i$, but  it may  also include charge vectors which do not belong to the list $(Q_1,\dots Q_r)$ if more than $r$ hyperplanes intersect at $p$, {\it i.e.} if the intersection is degenerate. In particular, 
$\kappa_{F,r}$ is the sum of all vectors in $\cQ_p$, irrespective of the flag $F$.  Note also that the matrix $\kappa_F$ is independent of the choice 
of representative $Q_F$ for the flag $F$. 

Now, let us promote the stability condition $\zeta\in \IR^K$ to a vector $\eta\in \IR^r$, by using the diagonal embedding $\eta_{a,s}=\zeta_a$, and perturbing slightly away from this point 
(the perturbation
must be chosen once for all, and be the same for all flags). A flag $F\in\cF_p$ is said to be stable  if $\eta$ belongs to the positive cone spanned by the vectors $\kappa_{F,i}$ with $i=1\dots r$, i.e.
\be
\label{etastab}
\eta=\sum_{i=1}^{r} \lambda_i \, \kappa_{F,i}\ ,\qquad \lambda_i>0
\ee 
In the
case where the vectors $\kappa_{i}(F)$ are linearly dependent, we discard the corresponding flag, since
they will not be stable for generic values of $\eta$. 
Let $\mathcal{F}_p(\eta)$ be 
the set of stable flags constructed from the list of hyperplanes $\cH_p$ meeting at $p$. 
The Jeffrey-Kirwan residue at $p$ is then the sum of iterated residues
\be
 \JKres_p \left(Z(u) , \eta\right)= \sum_{F\in \mathcal{F}_p(\eta)} 
 {\rm sign}(\det\kappa)\, 
 {\rm Res}_{\tilde u_r=0} \dots {\rm Res}_{\tilde u_1=0} \,  \tilde Z(\tilde u) 
\ee
where $\tilde u_i = H_{Q_i}(u)$ (or in matrix notation, $\tilde u=u\cdot Q_F$)
and  $\tilde Z(\tilde u) = Z(u) \,\de\tilde u/\de u$. 
The residue is most efficiently computed from the representation \eqref{Zv}, but one must of course
keep track of the Jacobians $\partial v/\partial u$ and $\partial u/\partial\tilde u$. 

\medskip

It is worth stressing that while the condition \eqref{etastab} restricts the possible flags contributing
to the index, it frequently happens that some of the allowed flags give a vanishing residue. In particular, note that in the absence of flavor fugacities and upon treating the hyperplanes 
$H_{ab,A,ss'}$, $A=1\dots \kappa_{ab}$ as a single hyperplane of degree $\kappa_{ab}$, the stability condition for  flags depends only on the sign of $\kappa_{ab}$, not on its absolute value, but 
the complex dimension $d$ of the moduli space $]\cM_Q$ does depend  on $|\kappa_{ab}|$, and the $\chi$-genus must vanish if $d<0$. It would be interesting to find a criterium predicting when a given
stable flag will produce a vanishing contribution.

\subsection{Residues and fixed points}
As explained in \cite{Hori:2014tda,Ohta:2014ria}, the residue formula \eqref{Omloc} arises
by applying supersymmetric localization to the functional integral defining the gauge theory 
on an Euclidean circle of radius $\beta$. For this, one constructs a linear combination $Q_{\hbar}$
of the 4 supercharges which squares to a combination $Q_{\hbar}^2=J$ of a $U(1)_R$ rotation and global flavor symmetries, such that the action becomes $Q_{\hbar}$-exact. Fixed points of $Q_{\hbar}$ are such that the complexified scalars $u_{a,s}$ in the Cartan part of vector multiplets become constant (independent of the coordinate $\tau$ along the thermal circle), while the
chiral multiplets must vanish for generic values of the $u_{a,s}$'s. The one-loop determinant 
for the off-diagonal components of the vector multiplets  and for the chiral multiplets gives a $(r,r)$
form $\omega$ on the space of the $u_{a,s}$'s, which has the topology of a cylinder $(\IC^\times)^r=(\IR \times S_1)^r$. By integration by parts, the integral can be rewritten as a sum of contour integrals around the poles at finite and infinite distance. At poles at finite distance in the $u$-plane, corresponding to an intersection of hyperplanes $H_1,\dots H_r$, the chiral multiplets (in case 
$H_i=H_{ab,A,ss'}$) or the off-diagonal scalars in the vector multiplets (in case $H_i=H_{a,s,s'}$)
may acquire a non-zero vev. These correspond to fixed points of the one-parameter subgroup  
of the flavor symmetry $\prod_{a,b, \kappa_{ab}>0} U(\kappa_{ab})$ acting
on  the Higgs
branch. The condition \eqref{etastab}  ensures that  fixed points with 
$\phi_{ab,A,ss'}\neq 0$ for $a,b,s,s'$ associated to the hyperplanes $H_i$ are allowed by 
 the D-term constraints \eqref{Dterm}. Moreover, the contribution of each fixed point agrees
 with the  Atiyah-Bott Lefschetz fixed point theorem  \cite{atiyah1968lefschetz} 
 for a K\"ahler manifold $X$ with a holomorphic
 action of $f:X\to X$, which lifts to an action $f^*$ on its cohomology $H^*(X)$, 
\be
\label{AtiyahBott}
 \sum_{p,q\geq 0} (-1)^{p+q-d} y^{2p-d} \, \Tr_{H^{p,q}} f^*
 =
\sum_{p\geq 0}
(-1)^{p-d} y^{2p-d} 
 \sum_{f(P)=P} 
  \frac{\Tr (\lambda^p \de f_P)}{\det_\IC (1-\de f_P)}
\ee
where $\de f_P$ is the linear action of $f$ on the holomorphic tangent space at the fixed point $P$, and 
$\lambda^p \de f_P$ is the $p$-th antisymmetric power of this map. In our context, $f$ is the $U(1)$ flavor symmetry determined by the chemical potentials $\theta_{ab,A}$. 
In Section \ref{sec_ab} below we discuss the relation between
 poles and fixed points in more detail in the case of Abelian quivers.

\subsection{Index in attractor chamber}
For any quiver, there is a special choice of stability conditions known as the attractor chamber \cite{Alexandrov:2018iao}
\be
\label{zetatt}
\zeta^*_a = - \sum_b \, \kappa_{ab} N_b \ .
\ee
This chamber precludes the existence of two-centered bound states,  since for any splitting
$\gamma=\gamma_L+\gamma_R$ of the dimension vector, or equivalently
$N_a=N_a^L+N_a^R$, the DSZ product $\langle \gamma_L,\gamma_R\rangle = \kappa_{ab} N_L^a N_R^b$ and the effective stability parameter $\zeta_L= N_L^a \zeta_a^\star = - N_R^a \zeta_a^\star$
have opposite sign. 
For a quiver without oriented loops, the index in this chamber automatically vanishes. This is because
such a quiver always admits (possibly more than) one sink, where all arrows are incoming, and (possibly more than) one source, where all arrows are outgoing. The D-term conditions \eqref{Dterm} have no solutions unless $\zeta_a<0$ for a sink, and $\zeta_a>0$ for a source. In contrast, at the attractor 
point $\zeta_a^*>0$ for sink, and $\zeta_a^*<0$ for a source. By the same token, the index vanishes
at the attractor point for any quiver which admits a sink or source, whether or not it contains an oriented loop; more generally the index vanishes
for any stability condition of the form $\zeta_a = - \mu_a \sum_b \, \kappa_{ab} N_b$ such that $\mu_a>0$ provided $\sum_a N_a \zeta_a = 0$. 
 
It is easy to see that the JK residue prescription is consistent with this vanishing 
property. Indeed, summing up
the relation \eqref{etastab} over the indices $s=1\dots N_a$ for each $a$, we get 
\be
\zeta= \sum_{i=1}^{r} \lambda_i\, \tilde\kappa_{i}
\ee
where $\tilde\kappa_{i}\in \IR^{K}$ 
gets contributions only from chiral multiplet charge vectors in $\cQ_p$. If $a$ 
is a
sink, all the contributions to the $a$-th component of $\tilde\kappa_{i}$ are strictly negative, 
so there are no stable flags when $\zeta_a>0$, in particular at the attractor point 
 $\zeta^*$. Similarly, if $a$ is a source, sink, all the contributions to the $a$-th component of $\tilde\kappa_{i}$ are strictly positive,  so there are again no stable flags when $\zeta_a<0$, in particular at the attractor point $\zeta^*$. 

In the absence of sources or sinks, in particular for quivers with oriented loops, the index at the attractor point does not necessarily vanishes, but rather gets contributions  from 
single-centered invariants and from scaling solutions thereof, as we discuss further in \S\ref{sec_abloop}.

\subsection{Cauchy-Bose formula}

One of the bottlenecks in the practical evaluation of computation of \eqref{Omloc} is the enumeration of intersections points and flags. Each non-Abelian group contributes $\mathcal{O}(N_a^2)$ singular hyperplanes, while the rank grows only linearly in $N_a$, so the number of possible $\ell$-plets grows exponentially. One remedy is to use 
the Cauchy-Bose identity \eqref{CauchyBose} to rewrite the vector multiplet determinant in 
\eqref{Zuhbar} as 
\be
\prod_{ s,s'=1\dots N \atop 
s\neq s'} 
\frac{\sin[ \pi \hbar ( u_{s'}-u_{s})]}{\sin[\pi \hbar( u_{s}-u_{s'}-1) ]} = 
 \sum_{\sigma\in S_N}  \frac{\epsilon(\sigma)\, \sin(\pi \hbar)^N}
 {\prod_{s=1}^N \sin[\hbar(u_s- u_{\sigma(s)}+1)]} 
\ee
Equivalently, in terms of the exponentiated variables $v_s=e^{2\pi\I\hbar u_s}$, 
\be
\label{Cauchyvec}
\frac{1}{\prod_{s=1}^N v_s} \prod_{s,s'=1\dots N \atop
s\neq s'}  \frac{  (v_{s'}-v_{s})}{v_{s}/y-y\, v_{s'}} 
 = (1/y-y)^N\, 
 \sum_{\sigma\in S_N} \, \frac{\epsilon(\sigma)}{\prod_{s=1}^N (v_s/y- y \,v_{\sigma(s)})} 
\ee
Using the elliptic generalization of  \eqref{CauchyBose} due to Frobenius
\cite{frobenius1882uber}
\be
\label{CauchyBoseEll}
\det\frac{1}{\theta_1(\mu_i-\nu_j)} =  \frac{\prod_{i<j}\theta_1(\mu_i-\mu_j)\, \theta_1(\nu_j-\nu_i)}
{\prod_{i,j}\theta_1(\mu_i-\nu_j)}\ ,
\ee
we may similarly write the vector multiplet determinant for the elliptic genus  in \eqref{Zuhbar} as 
\be
\prod_{ s,s'=1\dots N \atop 
s\neq s'} 
\frac{\vartheta_1[ \pi \hbar ( u_{s'}-u_{s})]}{\vartheta_1[\pi \hbar( u_{s}-u_{s'}-1) ]} = 
 \sum_{\sigma\in S_N}  \frac{\epsilon(\sigma)\, \vartheta_1(\pi \hbar)^N}
 {\prod_{s=1}^N \vartheta_1[\hbar(u_s- u_{\sigma(s)}+1)]} 
\ee
In each of these formulae, the denominator now involves $N$ hyperplanes rather than $N(N-1)/2$, 
which drastically simplifies the classification of intersections flags, though it requires sifting through $N!$ permutations rather than a single product of many hyperplanes. Fortunately, all permutations in the same conjugacy class  (labelled by the partition $\lambda=\sum_{\ell} \ell n_\ell$) turn out to produce the same contribution\footnote{This fact is non-trivial since the generalized stability vector 
$\eta\in \IR^r$ breaks the $S_N$ symmetry. The integrand coming from different permutations with the same cycle shape can be mapped to the same integrand by relabelling the integration variables
$u_{s,a}$, at the expense of permuting the entries in the flag $F=(Q_1,\dots, Q_r)$ used for integrating each of them. 
For non-degenerate intersections, these permutations do not affect the residue, so the result
is the same for all  permutations with the same cycle shape. For degenerate intersections, the 
change of variables typically permutes the flags as well (not only the entries $Q_i$ in a given flag) 
and a more detailed analysis is required.}, so it suffices to pick one particular permutation for each partition, and multiply the result by $\prod_\ell  
\ell^{n_\ell} n_\ell!$, corresponding to the number of permutations in the same conjugacy class.
As we shall see in Section \ref{sec_nonab}, 
if one applies this trick for each of the gauge groups $U(N_a)$,
the resulting sum of multi-partitions coincides with the MPS formula, at least in the case of 
non-Abelian quivers without oriented loops.

\section{Abelian quivers \label{sec_ab}}

In this section, we apply the residue formula in the context of Abelian quivers with or without loops.
In the absence of oriented loops, we show that the stability condition \eqref{etastab}
on flags coincides with 
the condition for existence of fixed points satisfying the D-term equations.

\subsection{Abelian quivers without loops  \label{sec_abnoloop}}

To demonstrate this in a simple example, let us focus on Abelian quivers without oriented loops (but possibly with non-oriented ones). The D-term equations 
\eqref{Dterm} can be written as 
\be
\label{DtermAb}
\zeta = \sum_{a,b; \kappa_{ab}>0}\, |\phi_{ab}|^2 \, (e_a - e_b )\ , \qquad
 |\phi_{ab}|^2:=  \sum_{A=1}^{\kappa_{ab}} |\phi_{ab,A}|^2
\ee
For a quiver without (oriented nor unoriented) loop, the rank $r=K-1$ is equal to the total number of edges $E=\#\{(a,b\in \cA_Q, \kappa_{ab}>0\}$ (counted {\it without} multiplicity), so each hyperplane $H_i$ originates from a set of chiral fields $\phi_{i,A}\equiv \phi_{a(i)b(i),A}$, and vice-versa, and \eqref{DtermAb} becomes
\be
\zeta
= \sum_{i=1}^{r} |\phi_i|^2\, Q_i\ .
\ee
For the flag $F=(Q_1,\dots Q_r)$, one has $\kappa_{F,i} =\sum_{k=1}^i Q_i$ and it is easy to check that the solution to $\zeta=\sum_{i=1}^r \lambda_i \kappa_{F,i}$ is given by 
\be
\lambda_i = |\phi_i|^2 - |\phi_{i+1}|^2 \quad \mbox{for} \quad i<r\ ,\qquad \lambda_r = |\phi_r|^2
\ee
This gives a transparent interpretation of the parameters $\lambda_i$ in \eqref{etastab} as difference of vevs of chiral fields.  In particular, the flag $F$ is stable if the stability parameters allow for solutions of the D-term equations where the only non-vanishing chiral fields $\phi_i$ satisfy $|\phi_1|^2 > |\phi_2|^2 > \dots > |\phi_{r}|^2>0$. Different flags correspond to different ordering of these vevs. The associated residue then computes the contribution of the fixed points satisfying these constraints. 

\medskip

As a simple example, consider the Abelian `star' quiver with $K$ vertices $V_0, V_1, \dots V_{K-1}$,
with $a_i$ arrows from $V_i$ to $V_0$, with $a_i$ a non-zero integer of arbitrary sign. We denote this quiver by $Q=S_{a_1,\dots a_K}$:
 \begin{center}
\begin{tikzpicture}[inner sep=2mm,scale=3]
  \node (a) at ( 0,0) [circle,draw] {$V_0$};
  \node (b) at ( 0.5,0.7) [circle,draw] {$V_1$};
  \node (c)  at ( 1,0) [circle,draw] {$V_2$};
  \node (d)  at ( 0.5,-0-.7) [circle,draw] {$V_3$};
  \node (e)  at ( -0.7,0.5) [circle,draw] {$V_{K}$};
 \draw [->] (b) to node[auto,swap] {$a_1$} (a);
 \draw [->] (c) to node[auto,swap] {$a_2$} (a);
 \draw [->] (d) to node[auto,swap] {$a_3$} (a);
 \draw [->] (e) to node[auto,swap] {$a_{K}$} (a);
 \draw [dotted] (-0.5,0) arc [start angle=140, end angle=320, radius=0.4];
\end{tikzpicture}
\end{center}

The D-term conditions 
\be
\sign(a_i)\, \sum_{A=1}^{|a_i|}  |\phi_{i,A}|^2 = \zeta_i\ ,\qquad i=1\dots K-1
\ee
admit solutions only when $\sign(a_i)=\sign(\zeta_i)$, in which case the quiver moduli space reduces
to a product of projective spaces $\cM_Q=\prod_i \IP^{|a_i|-1}$. This agrees with the result from the 
residue formula
\be
\chi_Q(\{1,\zeta_a\},y) = 
 \sum_{p} \JKres_p \left[    \prod_{i=1}^{K-1}
\left( \frac{y^{\sign a_i} u_0 - y^{-\sign a_i} u_i}{u_i-u_0} \right)^{|a_i|} 
\frac{\de u_i}{(y-1/y)u_i} \right]\ .
\ee
In this case, 
the singular hyperplanes $u_i=u_0$ have a non-degenerate intersection and a single
stable flag contributes, which depends on the ordering of the $\zeta_i$'s. The  result however is independent of that ordering, since the integral factorizes into a product of residues in each variable 
$u_i$: 
\be
\label{chi1AbK}
\chi_Q(\{1,\zeta_a\},y) = \prod_{i=1}^{K-1} \left[ (-1)^{a_i-1}  \frac{y^{|a_i|} - y^{-|a_i|}}{y-1/y} \right]
\ee
 In the presence of chemical potentials $\theta_{i,A}$ for the global symmetry group 
$\prod_i U(|a_i|)$, the intersection at the origin splits into $\prod_i |a_i|$ intersection points,
corresponding to a choice of $A_i\in\{1,\dots, |a_i|\}$ for each arrow, producing
\be
\label{chi1AbKnu}
\chi_Q(\{1,\zeta_a\},y,\theta) =\prod_{i=1}^{K-1} 
\left[ \sum_{A_i=1}^{|a_i|} \prod_{A\neq A_i} \frac{\nu_{i,A}/y- \nu_{i,A_i}\, y}
{\nu_{i,A_i}-\nu_{i,A}} \right]
\ee
Each of these contributions correspond to one fixed point with $\phi_{i,A}
=\sqrt{|\zeta_i|} \delta_{A,A_i}$, up to $U(1)$ gauge rotations of the phase of $\phi_{i,A}$.
The determinant $\det_\IC (1-\de f_P)$ 
appearing in the Lefschetz fixed point formula \eqref{AtiyahBott} evaluates to 
$\prod_{i} \prod_{A\neq A_i} (1- \nu_{i,A}/\nu_{j,A_i})$, in agreement with the denominator
in \eqref{chi1AbKnu}.
After summing over $A\neq A_i$,  each bracket in \eqref{chi1AbKnu} reduces to the corresponding bracket in \eqref{chi1AbK}, as expected since each factor $\IP^{|a_i|-1}$ is compact.
Moreover, as $y\to 1$, each of these non-degenerate intersections contributes $\pm 1$,
so that  $\chi_Q(\{1,\zeta_a\},y)$ counts the number of fixed points, up to an overall sign.

\subsection{Abelian quivers with unoriented loops  \label{sec_abunloop}}

For Abelian quivers with $h$ unoriented loops, the rank $r=K-1$ is equal to $E-h$,  where $E$ is the total of number of edges (not counting multiplicity). Choosing R-charges $R_{ab}$ and flavor fugacities $\theta_{ab,A}$ such that only non-degenerate intersections of $r$ hyperplanes 
$(Q_1,\dots Q_r)$ contribute, then only the chiral fields corresponding to $Q_1,\dots Q_r$ can get
non-trivial expectation values at the intersection, while the remaining $E-r=h$  must be set to zero. Removing these arrows from the original quiver $Q$ defines a `reduced' quiver $\widetilde{Q}$. In order for 
the matrix $\kappa_F$ to be non-degenerate, it is easy to see that $\widetilde{Q}$ must be a tree, which is therefore a spanning tree of the original quiver $Q$. The same argument as in the previous section shows that the stable flags are again in one-to-one correspondence with fixed points satisfying
the D-term equations, for any ordering of the vevs $|\phi_i|^2$ along the spanning tree. 

\subsubsection{Three node quiver}

As an example with one unoriented loop, 
let us consider a quiver with three nodes $V_1, V_2, V_3$ and 
 $a=\gamma_{12}<0,b=\gamma_{23}>0, c=\gamma_{31}<0$, which we denote by  
 $Q=C_{a,b,c}$:
 \begin{center}
\begin{tikzpicture}[inner sep=2mm,scale=2]
  \node (a) at ( 0,1.7) [circle,draw] {$V_1$};
  \node (b) at ( 1,0) [circle,draw] {$V_2$};
  \node (c)  at ( -1,0) [circle,draw] {$V_3$};
 \draw [->] (b) to node[auto,swap] {$|a|$} (a);
 \draw [->] (b) to node[auto] {$b$} (c);
 \draw [->] (a) to node[auto,swap] {$|c|$} (c);
\end{tikzpicture}
\end{center}
 The D-term equations 
\be
|\phi_{13}|^2 - |\phi_{21}|^2 =\zeta_1, \quad |\phi_{23}|^2 + |\phi_{21}|^2 =\zeta_2, \quad 
-|\phi_{13}|^2 - |\phi_{23}|^2 =\zeta_3
\ee
admit solutions only when $\zeta_2>0$ and $\zeta_3<0$. 
If $\zeta_1>0$, then $\phi_{13}\neq 0$ defines a point in $\IP^{|c|-1}$, fibered over 
$\IP^{|a|+b-1}$ parametrized by $(\phi_{21},\phi_{23})\neq (0,0)$, with index
\be
\label{chi111p}
\chi_Q^+(y) =(-1)^{a+b+c} \frac{(y^{|c|}-y^{-|c|}) (y^{|a|+b}-y^{-|a|-b})}{(y-1/y)^2} 
\ee
If instead $\zeta_1<0$,
then $\phi_{21}\neq 0$ defines a point in $\IP^{|a|-1}$, fibered over $\IP^{|c|+b-1}$ parametrized by $(\phi_{13},\phi_{23})\neq (0,0)$, with index
\be
\label{chi111m}
\chi_Q^-(y)  =(-1)^{a+b+c}  \frac{(y^{|a|}-y^{-|a|})(y^{|c|+b}-y^{-|c|-b)}}{(y-1/y)^2} 
\ee
The difference 
\be
\chi_Q^+(y) - \chi_Q^-(y) = (-1)^{a+c-1} 
\frac{y^{|c|-a}-y^{-|c|+a}}{y-1/y} \times (-1)^{b-1}  \frac{y^b-y^{-b}}{y-1/y} 
\ee
is interpreted as the contribution of a two-particle bound state with charges $\{\gamma_{1+2},\gamma_3\}$  \cite{Denef:2007vg}.

\medskip

Let us see how these results can be recovered from the localization formula \eqref{Omloc}.
For vanishing  R-charges or flavor fugacities,  the 
index is given by a residue of 
\be
Z_Q= 
\left( \frac{y u_1-u_2/y}{u_2-u_1}\right)^{|a|}\,
\left(\frac{y u_3-u_2/y}{u_2-u_3}\right)^{|b|}\,
\left(\frac{y u_3-u_1/y}{u_1-u_3}\right)^{c}\,
\widetilde{\prod_{i=1\dots 3}}\frac{\de u_i}{u_i(y-1/y)}
\ee
at the degenerate intersection $u_1=u_2=u_3$. The flag $(Q_{23},Q_{12})$, or equivalently 
$(Q_{23},Q_{13})$, has a degenerate 
 matrix $\kappa_F$ so does not contribute for generic values of the $\zeta_i$'s. 
For $\zeta_1<0,\zeta_2>0,\zeta_3<0$, the only stable flag is
$(Q_{12},Q_{23})\sim (Q_{12},Q_{13})$, reproducing \eqref{chi111m}. For $\zeta_1>0,\zeta_2>0,\zeta_3<0$, the only stable flag is instead
$(Q_{13},Q_{12})\sim (Q_{13},Q_{23})$, reproducing \eqref{chi111p}.  For generic
R-charges $R_{21}, R_{23}, R_{13}$, the degenerate intersection at $(0,0,0)$ splits into three
non-degenerate intersections $p_1=H_{31}\cap H_{12},
p_2=H_{12}\cap H_{23}, p_3=H_{23}\cap H_{31}$. We denote by $p_2^+=(Q_{12}, Q_{23})$
and $p_2^-=(Q_{23}, Q_{12})$ the two possible flags at $p_2$, which are no longer equivalent,
 and similarly for $p_3^\pm$, $p_1^\pm$. Depending on $\zeta_1$, there are now two stable flags,
 given in the following table,
 \be
 \begin{array}{|c|c|c|c|}
 \hline
 \zeta_1<\zeta_3<0 & \zeta_3<\zeta_1<0 & 0<\zeta_1<\zeta_2 & 0< \zeta_2<\zeta_1\\ \hline 
\{p_1^-, p_2^+\} & \{p_1^-, p_2^-\} & \{p_1^+, p_3^+\} & \{p_1^+, p_3^-\}\\ \hline
\end{array}
\ee
 In each interval, the contribution of each flag is a complicated rational function of $y^{\pm R_{ij}}$, but they combine in the same result \eqref{chi111p} or \eqref{chi111m}, depending on the sign of $\zeta_1$.
For generic flavor fugacities, the intersections become non-degenerate, and are in one-to-one correspondence with the possible fixed points. For those, only 2 out of the 3 set of arrows can be non-vanishing, leading to a total of either $|c|(|a|+b)$ or $|a|(b+|c|)$ fixed points, depending on 
the sign of $\zeta_1$. This agrees with \eqref{chi111p} and \eqref{chi111m} in the limit $y\to 1$.

\subsubsection{Four node quiver}
Next, we consider an example will become
relevant in \S\ref{sec_star}: an Abelian quiver with 4 nodes $V_{1,2,3,4}$ and 
arrows $\kappa_{12}=a>0, \kappa_{23}=b<0, \kappa_{34}=c>0, \kappa_{41}=d<0$, which we denote by $Q=C_{a,b,c,d}$:
 \begin{center}
\begin{tikzpicture}[inner sep=2mm,scale=2]
  \node (a) at ( -1,0) [circle,draw] {$V_1$};
  \node (b) at (0,1) [circle,draw] {$V_2$};
  \node (c)  at ( 1,0) [circle,draw] {$V_3$};
   \node (d)  at (0,-1) [circle,draw] {$V_4$};
 \draw [->] (a) to node[auto] {$a$} (b);
 \draw [->] (c) to node[auto,swap] {$|b|$} (b);
 \draw [->] (c) to node[auto] {$c$} (d);
  \draw [->] (a) to node[auto,swap] {$|d|$} (d);
\end{tikzpicture}
\end{center}
The D-term
equations 
\be
|\phi_{12}|^2 + |\phi_{14}|^2 =\zeta_1, \quad 
-|\phi_{12}|^2 - |\phi_{32}|^2 =\zeta_2,\quad 
|\phi_{32}|^2 + |\phi_{34}|^2 =\zeta_3, \quad 
-|\phi_{14}|^2 - |\phi_{34}|^2 =\zeta_4
\ee
admit solutions only when $\zeta_1,\zeta_3>0$ and $\zeta_2,\zeta_4<0$. Using the symmetries 
exchanging $(V_1,V_3)$ and $(V_2,V_4)$, there is no loss of generality in assuming that $c_1<c_3$, $c_2<c_4$. For vanishing R-charges and flavor fugacities,  the 
index is given by a residue of 
\be
\begin{split}
Z_Q=
\left(\frac{y u_2-u_1/y}{u_1-u_2}\right)^{a}\,
\left(\frac{y u_2-u_3/y}{u_3-u_2}\right)^{|b|}\,
\left(\frac{y u_4-u_3/y}{u_3-u_4}\right)^{c}\,
\left(\frac{y u_4-u_1/y}{u_1-u_4}\right)^{|d|}\,
\widetilde{\prod_{i=1\dots 4}}\frac{\de u_i}{u_i(y-1/y)}
\end{split}
\ee
at the degenerate intersection $u_1=u_2=u_3=u_4$. For $\zeta_2+\zeta_3>0$,
 we find a single stable flag $(Q_{32}, Q_{34}, Q_{12})$ contributing 
 \be
 \label{chi1111p}
 \chi_Q^+ = (-1)^{a+b+c+d+1} \, \frac{(y^{|b|}-y^{-|b|})(y^c-y^{-c})(y^{a+|d|} -y^{-a-|d|})}{(y-1/y)^3}
 \ee
 If instead  $\zeta_2+\zeta_3<0$, there is  a single stable flag $(Q_{32}, Q_{12}, Q_{34})$ contributing 
  \be
 \chi_Q^- = (-1)^{a+b+c+d+1} \, \frac{(y^a-y^{-a})(y^{|b|}-y^{-|b|})(y^{c+|d|} -y^{-c-|d|})}{(y-1/y)^3}
 \ee
The difference between the two contributions 
\be
 \label{chi1111m}
\chi_Q^+ - \chi_Q^- = (-1)^{a+b+c+d} \,
\frac{(y^{a-c} -y^{c-a})(y^{|b|}-y^{-|b|})(y^{|d|}-y^{-|d|})}{(y-1/y)^3}
\ee
is recognized as the index of a bound state of $\gamma_L=\gamma_1+\gamma_4$ and 
$\gamma_R=\gamma_2+\gamma_3$. In the presence of generic R-charges, the degenerate
intersection splits into 4 non-degenerate intersections, with 2 stable flags contributing in any chamber of the 
$(\zeta_2,\zeta_3)$ plane (for fixed values of $\zeta_1,\zeta_4$ satisfying the previous assumptions).
For generic flavor fugacities (and vanishing R-charges), the  degenerate
intersection splits into $a|b|c+a|bd|+ac|d|+c|bd|$ non-degenerate intersections, grouped in 4 subsets
corresponding to the 4 possible spanning trees. Depending on the stability conditions, only two of these subsets support a stable flag contributing $\pm 1$ in the limit $y\to 1$, in agreement with the 
previous answers  \eqref{chi1111p}, \eqref{chi1111m}.

%

\subsection{Abelian quivers with oriented loops  \label{sec_abloop}}
In the presence of oriented loops and absence of a superpotential, the
the quiver moduli space $\cM_Q$ is in general non-compact. 
If one naively tries to apply the residue formula \eqref{Omloc} 
for vanishing R-charges  and flavor fugacities,  the intersections are typically degenerate and non-projective. One way to resolve this problem is to switch on generic flavor fugacities $\theta_A$ for the chiral fields;  the resulting equivariant index is then a rational function of $y$ and 
$\nu_A=e^{\I\hbar\theta_A}$, with no natural way of taking the limit $\theta_A\to 0$. 
Another way out is  to allow for a generic superpotential, by tuning the R-charges $R_{ab}$ such that the gauge invariant product of chiral fields around any oriented loops carries R-charge two. Additionally, one may switch on some flavor fugacities so as to restrict the form of the superpotential, but possibly at the cost of opening non-compact directions.   

\medskip

Let us demonstrate this in the case of an Abelian  quiver 
$Q=C_{a,b,c}$ with 3 nodes $V_1, V_2, V_3$ 
with an oriented loop, $a=\gamma_{12}>0,b=\gamma_{23}>0, c=\gamma_{31}>0$:
 \begin{center}
\begin{tikzpicture}[inner sep=2mm,scale=2]
  \node (a) at ( 0,1.7) [circle,draw] {$V_1$};
  \node (b) at ( 1,0) [circle,draw] {$V_2$};
  \node (c)  at ( -1,0) [circle,draw] {$V_3$};
 \draw [->] (a) to node[auto] {$a$} (b);
 \draw [->] (b) to node[auto,swap] {$b$} (c);
 \draw [->] (c) to node[auto] {$c$} (a);
\end{tikzpicture}
\end{center}

Choosing $R_{12}=R_{23}=R_{31}=2/3$ so that the total R-charge of the loop is $2$, the 
index is given by a sum of residues of 
\be
Z_Q
=
\left( \frac{y^{1/3} u_2-u_1/y }{u_1-y^{-2/3} u_2}\right)^a  
\left( \frac{y^{1/3} u_3- u_2/y }{u_2-y^{-2/3} u_3}\right)^b
\left( \frac{y^{1/3} u_1- u_3/y }{u_3-y^{-2/3} u_1}\right)^c
\widetilde{\prod_{i=1\dots 3}}\frac{\de u_i}{u_i(y-1/y)}
\ee
Denoting $H_{12}=u_1-y^{-2/3} u_2$,  $H_{23}=u_2-y^{-2/3} u_3$,  $H_{31}=u_3-y^{-2/3} u_1$ 
the singular hyperplanes, and $Q_{12}, Q_{23}, Q_{31}$ the corresponding charge vectors, 
we find three non-degenerate intersections
$p_2=H_{12}\cap H_{23}, p_3=H_{23}\cap H_{31}, p_1=H_{31}\cap H_{12}$. We denote by $p_2^+$ 
 the flag $(Q_{12}, Q_{23})$, and by $p_2^-$ the flag $(Q_{23}, Q_{12})$,
 and similarly for $p_3^\pm$, $p_1^\pm$. A single 
 flag contributes for any signs of $\zeta_1,\zeta_2,\zeta_3$, given in the following table:
 \be
 \begin{array}{|c|c|c|c|}
 \zeta_1 & \zeta_2 & \zeta_3 & p \\ \hline
 - & - & + & p_1^+ \\
 + & - & + & p_1^- \\
  + & - & - & p_2^+ \\
 + & + & - & p_2^- \\
  - & + & - & p_3^+ \\
 - & +& + & p_3^- 
 \end{array}
 \ee
The  corresponding residue of course gives the same result as the computation based on the identification of the quiver moduli space as a complete intersection in a product of projective spaces \cite{Bena:2012hf,Manschot:2012rx}, since both rely on the same index theorem. 

  Across the wall at $\zeta_3=0$ (say), assuming $\zeta_1>0, \zeta_2<0$, the flag $p_1^-$ contributes
 when $\zeta_3>0$, while $p_2^+$ contributes when $\zeta_3<0$. Defining $Z_1(u_2,u_3)= {\rm Res}_{u_1=y^{-2/3} u_2}  Z_Q$, the difference of indices across the wall gives 
 \be
 \begin{split}
\Delta\chi= {\rm Res}_{u_2=y^{4/3} u_3} 
Z_1  + 
  {\rm Res}_{u_2=y^{-2/3} u_3} 
  Z_1 =  
  -\left(  {\rm Res}_{u_2=0}+  {\rm Res}_{u_2=\infty} \right)  Z_1 \ .
 \end{split}
 \ee 
The integrand $Z_1(u_2,u_3)$ is of the form $f(u_2,u_3) \frac{\de u_2}{u_2}$, where 
$f(u_2,u_3)$ is a rational function of degree 0. By homogeneity, the residues at $u_2=0$ and $u_2=\infty$ can be
traded for residues at $u_3=\infty$ and $u_3=0$, respectively. The latter are easily computed from
the limits
\be
\begin{split}
f(u_2,u_3) \stackrel{u_3\to 0}{\rightarrow} & \frac{(-y)^{b-c}}{(y-1/y)^2} \oint_{y^{-2/3} u_2} 
\frac{\de u_1}{u_1}
\left( \frac{y^{1/3} u_2-u_1/y }{u_1-y^{-2/3} u_2}\right)^a 
=  (-1)^{a+b+c} \frac{y^{b-c}(y^a-y^{-a})}{(y-1/y)^2} 
\\
f(u_2,u_3) \stackrel{u_3\to \infty}{\rightarrow} & \frac{(-y)^{c-b}}{(y-1/y)^2} \oint_{y^{-2/3} u_2} 
\frac{\de u_1}{u_1}
\left( \frac{y^{1/3} u_2-u_1/y }{u_1-y^{-2/3} u_2}\right)^a 
=  (-1)^{a+b+c} \frac{y^{c-b}(y^a-y^{-a})}{(y-1/y)^2} 
\end{split}
\ee
leading to 
\be
\Delta\chi=   (-1)^{b+c-1} \frac{y^{c-b}-y^{b-c}}{y-1/y} \times (-1)^{a-1}\frac{y^a-y^{-a}}{y-1/y} 
\ee 
 which is recognized as the contribution from the bound state with charges $\{\gamma_1+\gamma_2,\gamma_3\}$. This computation illustrates how discontinuities across the wall come
 from poles at infinity \cite{Hori:2014tda}. 
 
 \medskip
 
Let us now consider the attractor point \eqref{zetatt} where $\zeta_1^*=c-a, \zeta_2^*=a-b, \zeta_3^*=b-c$. For 
 $a,b<c$, the contribution comes from the flag $p_2^+$ if $b>a$, or from the flag $p_2^-$ if $a<b$.  The residue associated to $p_i^+$ and $p_i^-$ coincide, since there are only two hyperplanes intersecting at $p_i$. 
 The result should be compared with
 the Coulomb branch formula evaluated at the attractor point, 
 \be
 \label{Om111st}
 \Omega_*(1,1,1)=\OmS(1,1,1) + g_C(\{\gamma_1,\gamma_2,\gamma_3\}) + H(\{\gamma_1,\gamma_2,\gamma_3\},\{1,1,1\})
 \ee
 where $\OmS(1,1,1)$ is  the single-centered invariant, and  the modified Coulomb index is given 
 by \cite[\S 3.3]{Manschot:2012rx}
 \be
 \label{gC111}
 g_C(\{\gamma_1,\gamma_2,\gamma_3\}) = (-1)^k \frac{y^k+y^{-k}}{(y-1/y)^2}
 \ee
 \be
 H(\{\gamma_1,\gamma_2,\gamma_3\},\{1,1,1\}) =
 \begin{cases} -2(y-1/y)^{-2} & k \ {\rm even} \\
 (y+1/y) (y-1/y)^{-2} &  k \ {\rm odd}
\end{cases}
 \ee
 with $k=c-a-b$ (more generally, $k$ is equal to the largest of $a,b,c$ minus the sum of the other two). 
 Evaluating the l.h.s. of \eqref{Om111st} via the residue formula allows to read off the single-centered
 invariant $\OmS(1,1,1)$. Note that both  $\OmS(1,1,1)$ and $g_C+H$ vanish unless $c\leq a+b-2$,
 although $g_C$ by itself does not vanish for $a+b-2<c\leq a+b-c$. This is consistent with the fact
 that the quiver moduli space, when non-empty, has complex dimension $d=a+b-c-2$.
  
 \medskip
  
 It is instructive to compare the index  in the presence of a generic superpotential, computed
 using an assignment of R-charges such that the oriented loop carries charge 2, to the index
 for vanishing superpotential. As explained in \cite[\S2.5]{Manschot:2012rx}, the latter can be computed using the Harder-Narasimhan recursion, or equivalently Reineke's formula. Either way,
 one finds, in the chamber $\zeta_1>\zeta_2>0, \zeta_3<0$,
 \be
 \label{chiRein}
\chi^{W=0}_Q=(-1)^{a+b+c}y^{2-a-b-c}(1-y^{2a})(1-y^{2b})/(1-y^2)^2
\ee
which in particular is not invariant under $y\to 1/y$. This is consistent with the fact that the quiver moduli space with vanishing superpotential is a $\IP^{a-1}\times \IP^{b-1}$ bundle over the non-compact base $\IC^c$ \cite[\S 3.3]{Manschot:2012rx}. The effect of the superpotential is to restrict
to the point at the origin in $\IC^c$, and to a complete intersection of $c$ hypersurfaces in the fiber
over that point. 
The first operation removes a factor $y^{-c}$ in the index, and the second, by virtue of the 
Lefschetz hyperplane theorem, multiplies by an additional factor of $y^c$, as far as only negative powers of $y$ are concerned. The result of these two operations, $(-1)^{a+b+c}y^{2+c-a-b} (1-y^{2a})(1-y^{2b})/(1-y^2)^2$ agrees with the Coulomb index \eqref{gC111} up to positive powers of $y$, 
which therefore correctly captures the non-middle part of the cohomology, while the middle part
is captured by  the single-centered
 invariant $\OmS(1,1,1)$.

\medskip

Here we note that the same result \eqref{chiRein} can be obtained by applying the  formula \eqref{Omloc} for vanishing R-charge but generic flavor potentials $\nu_{1A}$ ($A=1,\dots a$),  $\nu_{2B}$ ($B=1,\dots b$), $\nu_{3C}$ ($C=1,\dots c$), which requires
extracting a suitable residue of   
\be
Z_Q=
\prod_{A=1}^a\left(\frac{u_2 y \nu_{1 A}-u_1/y}{u_1-u_2 \nu_{1 A}}\right)\,
\prod_{B=1}^b\left(\frac{u_3 y \nu_{2 B}-u_2/y}{u_2-u_3 \nu_{2 B}}\right)\,
\prod_{C=1}^c\left(\frac{u_1 y \nu_{3 C}-u_3/y}{u_3-u_1 \nu_{3 C}}\right)\,
\widetilde{\prod_{i=1\dots 3}}\frac{\de u_i}{u_i(y-1/y)}
\ee
Denoting $H_{12, A}=u_1-u_2\nu_{1 A}$, etc the singular hyperplanes,there are
$a \times b$ non-degenerate intersections $H_{12, A_0}\cap H_{23, B_0}$, $b \times c$ intersections $H_{2 B_0}\cap H_{31, C_0}$ and $a \times c$ intersections $H_{12, A_0}\cap H_{31, C_0}$. In the chamber $c_1>0,c_2<0,c_3<0$, only the flags $(Q_{23,B_0},Q_{12,A_0})$ contribute, each of them giving
\be
\label{fix111}
\prod_{A\neq A_0}^a\left(\frac{y \nu_{1 A}-\nu_{1 A_0}/y}{\nu_{1 A_0}- \nu_{1 i}}\right)\,
\prod_{B\neq B_0}^b\left(\frac{y \nu_{2 B}-\nu_{2 B_0}/y}{\nu_{2 B_0}- \nu_{2 j}}\right)\,
\prod_{C=1}^c\left(\frac{ y \nu_{1 A_0}\nu_{2 B_0}\nu_{3 C}-1/y}{1-\nu_{1 A_0}\nu_{2 B_0} \nu_{3 C}}\right)\, \xrightarrow{y\longrightarrow1} (-1)^{a+b+c}
\ee
Summing over all choices of $A_0 B_0$ and taking the limit $y\to 1$, we obtain $ (-1)^{a+b+c} a b$
in agreement with \eqref{chiRein}. Keeping $y\neq 1$, we recognize in \eqref{fix111} the
contribution of the fixed point under the flavor rotation 
$\varphi_{i,A}\mapsto e^{\I \hbar\theta_{i,A}}$ given by $P:\phi_{12,A}=\delta_{A,A_0}, 
\phi_{23,B}=\delta_{B,B_0}, \phi_{31,C}=0$. The flavor rotation act on the tangent space 
by a compensating gauge rotation,
\be
\Phi_{1A} \mapsto e^{\I(\theta_{1A}+\phi_1-\phi_2)} \Phi_{1A}\ ,\quad
\Phi_{2B} \mapsto e^{\I(\theta_{2B}+\phi_2-\phi_3)} \Phi_{2B}\ ,\quad
\Phi_{3C} \mapsto e^{\I(\theta_{3C}+\phi_3-\phi_1)} \Phi_{3C}\ ,\quad
\ee
with $\theta_{1A_0}+\phi_1-\phi_2 = 0$ and $\theta_{2B_0}+\phi_2-\phi_3 = 0$, so that its determinant
reproduces the denominator in \eqref{fix111}, in agreement with \eqref{AtiyahBott},
\be
{\rm det}_\IC (1-\de f_P)= \prod_{A\neq A_0} (1-\nu_{1A}/\nu_{1A_0}) 
\prod_{B\neq B_0} (1-\nu_{2B}/\nu_{2B_0}) 
\prod_{C} (1-\nu_{3C} \nu_{1A_0} \nu_{2B_0}).
\ee 
Rescaling the fugacities as  $\nu_{1,A} \to t^A \nu_{1,A}$,  $\nu_{2,B} \to t^B \nu_{1,B}$, $\nu_{3,C} \to t^C \nu_{3,C}$and then take the limit $t\to \infty$, then each 
of the $ab$ contributions \eqref{fix111} produces a single power of $y$, which sum up to the same 
result \eqref{chiRein} as obtained from Reineke's formula.

\section{Non-Abelian quivers \label{sec_nonab}}

In this section, we finally turn to non-Abelian quivers, and demonstrate how the residue formula \eqref{Omloc} combined with the Cauchy-Bose identity \eqref{Cauchyvec} gives  a natural
decomposition of the index in terms according to partitions of the total dimension vector, 
in agreement with the Coulomb branch formula.

\subsection{Kronecker quiver with rank $(N,1)$ \label{sec_kronN1}}
We consider a quiver $Q=K_m$ with two nodes $V_1, V_2$ and $m>0$ arrows from  $V_1$ to $V_2$:
\begin{center}
\begin{tikzpicture}[inner sep=2mm,scale=2]
  \node (a) at ( -1,0) [circle,draw] [label=below:$\zeta_1$] {$N_1$};
  \node (b) at ( 1,0) [circle,draw]  [label=below:$\zeta_2$] {$N_2$};
 \draw [->] (b) to node[auto,swap] {$m$} (a);
 \end{tikzpicture}
\end{center}
We choose the  dimension vector $(N_1,N_2)=(N,1)$ and stability parameters $\zeta_1<0$, $\zeta_2>0$ such that $N \zeta_1+  \zeta_2=0$. For $m\geq N$, the quiver moduli space $\cM_Q$ is known to be the Grassmannian $G(N,m)$ of $N$-dimensional planes inside $\IC^m$, of dimension
$d=N(m-N)$, with $\chi$-genus given by
\be
\label{IBin}
\chi_Q(y)=\frac{(-y)^{-N(m-N)}\, [m,y]!}{[N,y]!\, [m-N,y]!}
\ee
where $[m,y]!=\prod_{k=1}^m (1-y^{2k}/(1-y^2)$ is the deformed factorial. For $m<N$, the moduli space is empty. Our aim is to rederive
this well-known result using localization, and explain how the decomposition predicted by the Coulomb branch formula naturally emerges in this context. 

\medskip

In order to apply the localization formula  \eqref{Omloc}, we upgrade the
stability vector $\zeta=(\zeta_1,\zeta_2)$ to $\eta=(\eta_1,\dots,\eta_N,\zeta_2)$ where 
$\eta_1<\eta_2<\dots<\eta_N$ and $\eta_s\simeq \zeta_1$ for $s=1 \dots N$. We denote by 
$(v_1,\dots,v_N)$ and $u$
the exponentiated Cartan variables associated to the vertices $V_1$ and $V_2$.  The  
equivariant $\chi$-genus
is then a sum of residues of 
\be
Z_Q = 
\left(\prod_{s\neq s'} \frac{u_{s'}-u_{s}}{u_s/y-y\, u_{s'}} \right)
\prod_{s=1}^N
\left( \prod_{A=1}^m  \frac{y \,\nu_A \, u_s- u/y}{u-\nu_A\, u_s}  \frac{\de u_s}{u_s(y-1/y)} \right)
\ee
As argued in \cite{Cordova:2015qka,Kim:2015oxa}, 
residues involving vector multiplet poles $u_s/y-y u_{s'}$ always vanish, so the only
contribution comes from the intersection of the hyperplanes $(u-\nu_{A_1} u_1,u-
\nu_{A_2}u_2,\dots u- \nu_{A_N}u_N)$, for any subset of $N$ distinct elements $\{A_1,\dots, A_N\}  \subset \{1,\dots, m\}$, 
taken in this particular order for the above choice of $\eta$. In this way one arrives 
at\footnote{This agrees with \cite[\S 4.6]{Benini:2013xpa}, upon taking the limit $\tau\to\I\infty$ in their formula}
\be
\chi_Q(y,\nu_A) = \frac{1}{N!} \oint Z_Q 
= \sum_{\cI \in C(N,m)} \prod_{A\in \cI} \prod_{B\notin \cI} \frac{\nu_A/y - \nu_B y}
{\nu_A-\nu_B}
\ee
where $\cI$ runs over all subsets of $N$ distinct elements $\{A_1,\dots, A_N\}  \subset \{1,\dots, m\}$. Each term in this sum originates from a $U(1)$ fixed point on $G(N,m)$, where the $N\times m$ matrix $\phi_{sA}$ has $N$ non-zero elements in positions $(s,A_s)$. The result can be shown to be independent\footnote{For this it suffices to show that the residue at the  potential singularities $\nu_A=\nu_B$ vanish.}  of the fugacities $\nu_A$, and equal to the deformed binomial coefficient \eqref{IBin}.

\medskip

In order to explain the connection with the Coulomb branch formula, we set to one the flavor 
fugacities $\nu_A$ and fix the gauge $u=1$, obtaining 
\be
\label{JKGr}
\chi_Q(y)=\frac{1}{(y-1/y)^N\, N!} \int \prod_{s=1}^N \frac{\de u_s}{2\pi\I \, u_s}
\prod_{s\neq s'} \frac{u_{s'}-u_{s}}{u_s/y-y\, u_{s'}} 
\prod_{s=1}^N \left(  \frac{y \, u_s- 1/y}{1-\, u_s}  \right)^m
\ee
where the integral runs over a product of small circles around $u_s=1$. 
Using the Cauchy-Bose  identity \eqref{Cauchyvec}, we can  rewrite \eqref{JKGr} as a sum over permutations,
\be
\label{JKGrP}
\chi_Q=\frac{(-1)^{N}}{N!} \int \prod_{s=1}^N \frac{\de u_s}{2\pi\I }
\sum_{\sigma\in S_N} \frac{ \epsilon(\sigma)}{\prod_{s=1}^N (u_s/y - y u_{\sigma(s)})}
\prod_{s=1}^N \left(  \frac{y \, u_s- 1/y}{1-\, u_s}  \right)^m
\ee
Decomposing each permutation $\sigma$ into a product of cycles $\prod_\ell (\cC_{\ell})^{n_\ell}$,  the integral factorizes into a product of factors associated to each cycle $\cC_\ell$ of length $\ell$, 
\be
\label{IcycleSum}
\chi_Q=\frac{(-1)^N}{N!} \sum_{\sigma\in S_N} \epsilon(\sigma)\,\prod_\ell (\cI_\ell)^{n_\ell}
\ee
where $\cI_\ell$ denotes the integral (with $x_i\equiv x_{i+L}$),
\be
\cI_\ell=\prod_{i=1}^\ell \oint_1 \frac{\de x_i}{2\pi\I }
 \frac{1}{\prod_{i=1}^\ell (y \, x_i - x_{i+1}/y)}
\left(\frac{y \, x_i -1/y}{x_i-1}\right)^m
\ee
Moreover, the signature of the permutation is $\epsilon(\sigma)=\prod_\ell (-1)^{n_\ell(\ell-1)}$.
By successively integrating over each $u_i$, one can establish that 
\be
\label{Icycle}
\cI_{\ell} = \frac{y^{m\ell}-y^{-m\ell}}{y^\ell-y^{-\ell}}
\ee
For this, one may deform the contour around $x_i=1$ to a sum of contours  around $x_i=x_{i+1}/y^2$ and $x_i= y^2 x_{i-1}$. The residue at $x_i=x_{i+1}/y^2$ is regular at $x_{i+1}=1$ (due to a cancellation of factors of $(x_{i+1}-1)^m$ in the nominator and denominator), so can be dropped.
The residue at $x_i=y^2 x_{i-1}$ produces a factor $[(y^2 x_{i-1}-1/y^{2})/(y x_{i-1}-1/y)]^m$, which cancels partially against $[(yx_{i-1}-1/y)/(x_{i-1}-1)]^m$, etc. Alternatively, one may reinstate the flavor fugacities and evaluate
\be
\cI_{\ell} =\int \prod_{i=1}^\ell \frac{\de x_i}{2\pi\I }
 \frac{1}{\prod_{i=1}^\ell (x_i/y - y x_{i+1})}
 \prod_{A=1}^m \frac{y \nu_A x_i   -1/y}{1-\nu_A x_i}
\ee
where the integral circles around each pole at $x_i=\nu_{A(i)}$ for all maps $A:[1,\ell]\to[1,m]$.
In the limit $y\to 1$ it is easy to check that constant maps contribute $+1$ while non-constant maps contribute 0, leading to $\I_\ell\to m$. 
For $y\neq 1$ but assuming $\nu_1\ll \dots \ll \nu_m$, one also finds that the only non-vanishing contribution come from constant maps $A(i)=A$, contributing $y^{\ell(2A-m-1)}$, in agreement with \eqref{Icycle}.

\medskip

Using  \eqref{Icycle}, and noting that the number of permutations with cycle shape $\lambda=\sum \ell n_\ell$
is $N!/ \prod_\ell {\ell^{n_\ell}\, n_\ell!}$, one may rewrite \eqref{IcycleSum} as a sum over partitions of $N$,
\be
\label{IpartSum}
\chi_Q=(-1)^{N(m+1)}
\sum_{N=\sum_{\ell=1}^N \ell n_\ell}  \prod_{\ell=1}^N \frac{[(-1)^{\ell+1}\cI_{\ell}]^{n_\ell}}{\ell^{n_\ell}\, n_\ell!}
=\sum_{N=\sum_{\ell=1}^N \ell n_\ell}  \prod_{\ell=1}^N \frac{[(-1)^{m\ell+1}\cI_{\ell}]^{n_\ell}}{\ell^{n_\ell}\, n_\ell!}
\ee
Note that this coincides with the cycle index $Z_N(\{t_\ell\})$ for the permutation group $S_N$, evaluated at $t_\ell=(-1)^{m\ell+1} \cI_\ell$. In order to evaluate the sum over partitions, it is  expedient to construct the generating function,
\be
G=\sum_{N=0}^{\infty} Z_N(\{t_\ell\}) \, t_0^N = \exp\left( -\sum_{\ell=1}^{\infty} \frac{(-y)^{m\ell}-(-y)^{-m\ell}}{\ell (y^\ell-y^{-\ell})}\, t_0^\ell \right)\ .
\ee
The result agrees with $Z_{\rm halo}$ in \cite[(4.72)]{Manschot:2010qz}, up to a change $m\to -m$ since we have $m>0$ while the result of loc.cit. assumed $m<0$.

\medskip

We can now compare \eqref{IpartSum} with the Coulomb branch formula \eqref{CoulombForm1} for this system (or equivalently \eqref{MPS}), 
\be
\chi_Q=\sum_{N=\sum_{\ell=1}^N \ell n_\ell}  
\frac{g_C( \{n_1\times \gamma_1, n_2\times 2\gamma_1, \dots n_\ell \times \ell\gamma_1,\gamma_2\})}
{\prod_{\ell=1}^N n_\ell!}
\prod_{\ell=1}^N [\overline{\Omega}_S(\ell\gamma_1)]^{n_\ell}\, \overline{\Omega}_S(\gamma_2)
\ee
where $n\times \ell\gamma$ denotes $n$ copies of the vector $\ell\gamma$. Since the only non-vanishing DSZ products are $\langle \ell\gamma_1,\gamma_2\rangle=m \ell$, 
the Coulomb index factorizes
into 
\be
g_C(n_1\times \gamma_1, n_2\times 2\gamma_1, \dots n_\ell \times \ell\gamma_1,\gamma_2) = \prod_{\ell=1}^N\, 
\left( \frac{y^{m\ell}-y^{-m\ell}}{y-1/y} \right)^{n_\ell}
\ee
which is also the $\chi$-genus of the Abelian quiver $Q(\{\gamma_i\})$ described below \eqref{MPS}. 
Moreover, in the absence of loops, the only non-vanishing single centered invariants are $\Omega_S(\gamma_1)=\Omega_S(\gamma_2)=1$, hence
\be
\overline{\Omega}_S(\ell\gamma_1) = \frac{y-1/y}{\ell(y^\ell-y^{-\ell})}\ ,\quad 
\overline{\Omega}_S(\ell\gamma_2) = 1 
\ee
Combining these relations, we conclude that  \eqref{IpartSum} is in perfect agreement with the Coulomb branch formula. Moreover, the sum over partitions of $\gamma=N\gamma_1+\gamma_2$
clearly originates from the sum over conjugacy classes in the permutation group  $S_N$.

\subsection{Star quivers \label{sec_star}}
We now turn to a generalization of both the Abelian star quiver $S_{\{a_i\}}$ considered in \S\ref{sec_abnoloop}, and the Kronecker quiver $K_m$ of rank $(N,1)$ in the previous subsection. Namely, we consider a quiver with $K+1$ vertices $V_0,V_1,\dots V_{K}$, with $a_i>0$ arrows from $V_i$ to $V_0$, and with dimension vector $(N,1,1,\dots)$. Our aim is to evaluate the index using the Cauchy-Bose formula for the contribution of the $U(N)$ vector multiplets, and show that the resulting decomposition agrees
with the Coulomb branch formula \eqref{CoulombForm1}, or equivalently the MPS formula \eqref{MPS}.  

Clearly, the moduli space is trivial unless $\zeta_0<0$ and $\zeta_i>0$ for $i=1\dots K$.
Up to relabelling the nodes, we can assume that $\zeta_1> \zeta_2>\dots > \zeta_{K}$.
We shall further assume that $\zeta_1\gg \zeta_2 \gg \dots \gg \zeta_{K}$, and upgrade
the stability vector to $\eta=(\eta_1,\dots,\eta_N,\zeta_1,\dots \zeta_{K} )$ with 
$\eta_1<\eta_2<\dots<\eta_N$ and $\eta_s\simeq \zeta_0$ for $s=1 \dots N$. We denote by 
$(v_1,\dots,v_N)$ and $(u_1,\dots u_{K})$
the exponentiated Cartan variables associated to the vertices $V_0$ and $V_i$.  The  $\chi$-genus
is then a sum of residues of 
\be
\label{ZQstar}
Z_{Q}= \frac{1}{(y-1/y)^{N+K-1}}  
\left( \prod_{s\neq s'} \frac{v_{s'}-v_{s}}{v_s/y-y\, v_{s'}} \right)
\left( \prod_{s=1\dots N \atop i=1\dots K}
 \left[  \frac{y \, v_s- u_i/y}{u_i- v_s} \right]^{a_i}  \right)
\prod_{i=1}^{K} \frac{\de u_i}{u_i} 
\widetilde{\prod_{s=1\dots N}}  \frac{\de v_s}{v_s} 
\ee
As for the Grassmannian in the previous subsection, residues involving vector multiplet poles 
$v_s/y-y\, v_{s'}$ always vanish  \cite{Cordova:2015qka,Kim:2015oxa}, 
so the only contribution comes from the intersection of
the $N K$ hyperplanes $u_i - v_s$, which is degenerate if $K\geq 2$. For the above choice of $\eta$,
this intersection carries a single stable flag $(u_1-v_1,~u_1-v_2,~\dots,~u_1-v_N,~u_2-v_N,~u_K-v_N)$, corresponding to the following integration prescription 
\be
\chi_Q = \frac{1}{N!} \oint Z_Q\ ,\quad \oint:=
(-1)^{N-1} \oint_{v_N} \frac{\de u_K}{u_K} \dots \oint_{v_N} \frac{\de u_1}{u_1}  \oint_{u_1} \frac{\de v_{N-1}}{v_{N-1}} \dots \oint_{u_1} \frac{\de v_1}{v_1}
\ee
We now apply the Cauchy-Bose formula \eqref{Cauchyvec} to the vector multiplet product
in \eqref{ZQstar}, and collect contributions according to the cycle shape of the permutation $
\sigma\in S_N$, corresponding to a partition $\lambda=\ell_1+\dots + \ell_p=\sum \ell n_\ell$,
where $n_\ell$ is the number of $n_i$'s equal to $\ell$ 
For convenience we relabel the Cartan variables 
for the non-Abelian group accordingly,
\be
\{v_1, \dots , v_N \} \rightarrow \{ v^{(1)}_1, \dots, v^{(1)}_{\ell_1};~v^{(2)}_1, \dots, v^{(2)}_{\ell_2};~\dots;~v^{(p)}_1, \dots, v^{(p)}_{\ell_p}\} \, .
\ee
All permutations with the same cycle shape $\lambda$ give the same contribution to the $\chi$-genus,
\be
\begin{split}
\chi_\lambda &= (-1)^{N-p} (y-y^{-1})^{-N-K+1} (-1)^{N-1} (1-y^2)^N \oint_{v_N} 
\frac{\de u_K}{u_K} \dots \oint_{v_N} \frac{\de u_1}{u_1} \oint_{u_1} \frac{\de v_{N-1}}{v_{N-1}} \dots \oint_{u_1} \frac{\de v_1}{v_1}
 \\
& \left[ \prod_{\beta=1}^p \frac{v^{(\beta)}_1 \dots v^{(\beta)}_{\ell_\beta} }{ (v^{(\beta)}_1 - y^2 v^{(\beta)}_2) \dots (v^{(\beta)}_{\ell_\beta} - y^2 v^{(\beta)}_1) }   \right]
\prod_{s=1}^N \prod_{\alpha=1}^K \left( \frac{y v_s - y^{-1} u_\alpha}{u_\alpha - v_s} \right)^{a_\alpha}  \, . \label{ResidueStar}
\end{split}
\ee
where $v_N$ can be gauged fixed to any value.
The sum over all  permutations with the same cycle shape turn the prefactor $1/N!$ into a factor
$1/( \prod_{\ell}n_\ell!  \ell^{n_\ell} )$. According to the MPS formula \eqref{MPS}, \eqref{ResidueStar}
should coincide with 
\be
\tilde\chi_\lambda =  \frac{(y-y^{-1})^p}{\prod_{\beta=1}^p (y^{\ell_\beta} - y^{- \ell_\beta})}
\chi \left(Q_\lambda \right) 
\ee
where $Q_\lambda$ is  the Abelian quiver 
\be
\nonumber
\begin{tikzcd}
1_1 \arrow[dd,shift right=.2ex,swap] \arrow[ddrr,shift right=.2ex,swap] \arrow[ddrrrr,shift right=.2ex] &
 \dots & \arrow[ddll,shift right=.2ex] 1_\alpha \arrow[dd,shift right=.2ex] \arrow[ddrr,shift right=.2ex] &
 \dots &
1_K \arrow[ddllll,shift right=.2ex] \arrow[ddll,shift right=.2ex]  \arrow[dd,shift left=.2ex] \\
&&&&\\
1_1 & \dots & 1_\beta & \dots & 1_p
\end{tikzcd} 
\ee
with $\ell_\beta \times  a_i$ arrows going from the node $1_i$ on the top row to the node $1_\beta$ on the bottom row. The $\chi$-index for $Q_\lambda$ follows from the residue formula \eqref{Omloc},
\be
\tilde\chi_\lambda =  
\frac{(-1)^{p-1} (y-y^{-1})^{-K+1}}{\prod_{\beta=1}^p (y^{\ell_\beta} - y^{- \ell_\beta})} \oint_{v_p} \frac{\de u_K}{u_K} \dots \oint_{v_p} \frac{\de u_1}{u_1} \oint_{u_1} \frac{\de v_{p-1}}{v_{p-1}} \dots \oint_{u_1} \frac{\de v_1}{v_1}  
 \prod_{r=1}^p \left[ \prod_{\alpha=1}^K \left( \frac{y v_r - y^{-1} u_\alpha}{u_\alpha -v_r} \right)^{a_\alpha} \right]^{\ell_r}  \, . 
 \label{MPSstar}
\ee
where $v_p$ can be gauge fixed to any value.
We shall now prove, by induction on $K$, that \eqref{ResidueStar} and \eqref{MPSstar} coincide.
For $K=2$, this is the Kronecker quiver with rank $(N,1)$ discussed in the previous subsection. We
shall assume that the equality $\chi_\lambda=\tilde\chi_\lambda$ holds for star quivers 
with $K-1$ nodes, and show that it continues to hold for $K$ nodes. 

\medskip

To show that \eqref{ResidueStar} and \eqref{MPSstar} coincide, we focus on the last contour integral
over $u_K$ around $v_N$ in , \eqref{ResidueStar}, or around $v_p$ in   \eqref{MPSstar}, which we free to identify by a choice of gauge. Denoting by $\chi_\Lambda(u_K,v)$ and 
$\tilde\chi_\Lambda(u_K,v)$ the two integrands, we need to prove 
\be
\oint_{v} \frac{\de u_K}{u_K} \left[ \chi_\Lambda(u_K,v)- \tilde\chi_\Lambda(u_K,v) \right]= 0 \, , \label{goal}
\ee
By construction, $\chi_\Lambda(u_K,v)$ and $\tilde\chi_\Lambda(u_K,v) $ are homogenous rational functions of degree 0. The recursion shows that their difference
can be put to the form
\be
\chi_\Lambda(u_K,v)- \tilde\chi_\Lambda(u_K,v)   = \frac{N_m(u_K, v)}{(u_K-v)^m} \, ,
\ee
where $m$ is a non-negative integer and $N_m(u_K,v)$ is a homogeneous polynomial of degree $m$. In particular,  the denominator does not have any factor of $u_K$ or $v$. The equality \eqref{goal} 
will follow if we can show that the numerator factorizes as 
\be
\label{Nfac}
N_m(u_K,v) = v^{m_1} u_K^{m_2} \tilde N_{m-m_1-m_2}
\ee
with $m_1,m_2\geq 1$. Indeed, setting $v=1$ by homogeneity, it follows from \eqref{Nfac} that
\be
\oint_{1} \frac{\de u_K}{u_K} \left[\chi_\Lambda(u_K,1)- \tilde\chi_\Lambda(u_K,1)  \right]
  = \oint_1 \de u_K \frac{ u_K^{m_2-1} \tilde{N}_{m-m_1-m_2}(u_K,1)}{(u_K-1)^m}  = 0 
\ee
since the numerator has monomials of degree less than $m-1$ are annihilated by the $(m-1)$ differentiations needed to extract the residue. To show that $N_m(u_K,v)$ factorizes as in \eqref{Nfac},
it suffices to check that $I_K(u_K,v)$ vanishes both when $u_K\to 0$ and when $u_K\to \infty$ (which is equivalent to $v\to 0$ by homogeneity). In either of those limits, the integrands 
$\chi_\lambda(u_K,v)$ and $\tilde\chi_\lambda(u_K,v)$ reduce to the indices for the star
quiver obtained by removing the node  $v_K$. By induction, the equality  $\chi_\lambda=\tilde\chi_\lambda$ therefore holds for star quivers with arbitrary number of Abelian nodes.

\subsection{Kronecker quiver with rank $(N_1,N_2)$}
We now consider the Kronecker quiver $K_m$  introduced in \S\ref{sec_kronN1}, now for general dimension vector.  The $\chi$-genus of the moduli space $\cM_Q$ has been computed in the mathematics literature \cite{weist2013localization,Mozgovoy:2012,ReinekeMPS} 
using the Atiyah-Bott Lefschetz fixed point theorem, and in the physics literature \cite{Ohta:2014ria,Cordova:2015qka,Kim:2015oxa,Cordova:2015zra} using supersymmetric localization.  
Our goal here is to clarify the relation between the two approaches, and explain how the Coulomb branch formula arises from applying the Cauchy-Bose formula to both nodes. 
\medskip

The quiver moduli space $\cM_Q$ is non-trivial in the chamber $\zeta_1<0<\zeta_2$ with
$N_1\zeta_0+N_2\zeta_1=0$, in which case its complex dimension is $d=m N_1 N_2 - N_1^2 - N_2^2+1$. In order to apply the residue formula \eqref{Omloc}, we upgrade the
stability vector $\zeta=(\zeta_0,\zeta_1)$ to $\eta=(\eta_1,\dots,\eta_{N_1},\tilde\eta_1,
\dots \eta_{N_2})$ where 
$\eta_1<\dots<\eta_{N_1}<0 <\tilde\eta_1 < 
\dots< \eta_{N_2}$ and $\eta_s\simeq \zeta_1,\tilde\eta_{s'}\simeq \zeta_2$ for $s=1 \dots N_1, s'=1 \dots N_2$. The equivariant $\chi$-genus is then given by a suitable residue of 
\be
\begin{split}
Z_Q = &\frac{1}{(y-1/y)^{N_1+N_2-1}}  
\prod_{s,s'=1\dots N_1\atop  s\neq s'} \frac{u_{s'}-u_{s}}{u_s/y-y\, u_{s'}} 
\prod_{s.s'=1\dots N_2\atop  s\neq s'} \frac{v_{s'}-v_{s}}{v_s/y-y\, v_{s'}}  \\
& \times 
\prod_{s=1\dots N_1\atop s'=1\dots N_2} 
 \prod_{A=1}^m  \frac{y \,\nu_A \, v_{s'}- u_{s}/y}{u_{s}-\nu_A\, v_{s'}} 
\prod_{s=1\dots N_1}  \frac{\de u_s}{u_s} 
\widetilde{\prod_{s'=1\dots N_2}}   \frac{\de v_{s'}}{v_{s'}} 
\end{split}
\ee
For $(N_1,N_2)$ coprime, it turns out that residues involving the vector multiplet hyperplanes $u_s/y-y\, u_{s'}, v_s/y-y\, v_{s'}$ vanish. For generic fugacities, the intersections of the chiral multiplet 
hyperplanes are non-degenerate, and in one-to-one correspondence with bipartite trees linking $N_1$ `black' vertices $e_1,\dots e_{N_1}$ to $N_2$ `white' vertices $f_1,\dots, f_{N_2}$, with 
edges $e_{s}-f_{s'}$ colored by an integer $A\in\{1,\dots, m\}$. Each edge  represents
one hyperplane $u_{s}-\nu_A\, v_{s'}$, and each tree corresponds to a particular non-degenerate intersection, which may or may not carry a stable flag. As usual, stable flags correspond 
to fixed points of the toric action $\phi_{Ass'}\mapsto \nu_A\, \phi_{Ass'}$ on the space of solutions
to the D-term equations \eqref{Dterm}, up to gauge transformations. Since the quiver moduli space is compact, the dependence on $\theta_A$ must cancel after summing over all  fixed points. In the limit $y\to 1$, each stable flag contributes a rational number $\pm 1/(N_1! N_2!)$, which must add up to the Euler 
characteristic which is integer. Alternatively, by assuming that the $\theta_A$'s are hierarchically
ordered, e.g. $\theta_1 \ll \theta_2 \ll \dots \ll \theta_m$, one may associate a given power 
$\pm y^{2J_3} /(N_1! N_2!)$ to each fixed point, which must add up to the $\chi$-genus.

\subsubsection{Rank $(2,3)$ \label{sec_k23}}

\begin{figure}[h]
\begin{center}
\begin{tikzpicture}[scale=1, minimum size=4mm]
\begin{scope}[shift={(-3,0)}]
  \node (u1) at ( -1,1) [circle,draw] {$u_1$};
  \node (u2) at ( -1,-1) [circle,draw] {$u_2$};
  \node (v1) at (1,1.5) [circle,draw] {$v_1$};
  \node (v2) at (1,0) [circle,draw] {$v_2$};
  \node (v3) at (1,-1.5) [circle,draw] {$v_3$};
  \node (c) at (0,-2.5) {$F_{ABCD}$} ;
  \draw (u1) to node[auto] {$A$}  (v1); 
  \draw (u1) to node[auto] {$B$}  (v2);
   \draw (u1) to node[auto,swap] {$C$}  (v3); 
  \draw (u2) tonode[auto,swap] {$D$}   (v3); 
\end{scope}
\begin{scope}[shift={(3,0)}]
  \node (u1) at ( -1,1) [circle,draw] {$u_1$};
  \node (u2) at ( -1,-1) [circle,draw] {$u_2$};
  \node (v1) at (1,1.5) [circle,draw] {$v_1$};
  \node (v2) at (1,0) [circle,draw] {$v_2$};
  \node (v3) at (1,-1.5) [circle,draw] {$v_3$};
  \node (c) at (0,-2.5) {$W_{ABCD}$} ;
  \draw (u1) to node[auto] {$A$}  (v1);
  \draw (u1) to node[auto] {$B$}  (v2); 
  \draw (u2) tonode[auto] {$C$}   (v2); 
  \draw (u2) to node[auto] {$D$}  (v3); 
\end{scope}
\end{tikzpicture}
\end{center}
\caption{Graphs associated to the flags contributing to the equivariant $\chi$-genus of the 
Kronecker quiver $K_m(2,3)$ (up to permutations of $\{u_s\}$ and $\{v_{s'}\}$)
\label{fig:flags23}}
\end{figure}
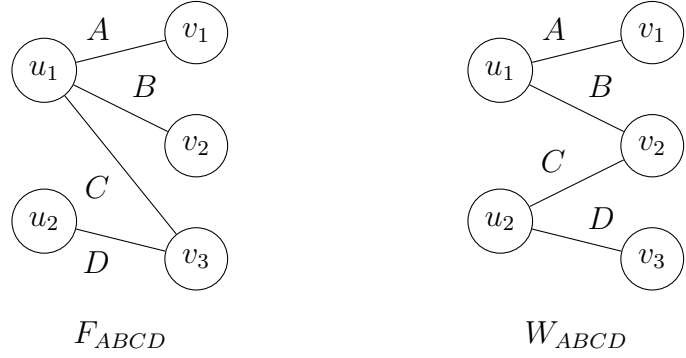

For illustration, we consider the rank $(2,3)$ case in some detail. For generic values of the fugacities, one finds contributions from two types of graphs, depicted in Figure \ref{fig:flags23}: 
\begin{itemize}
\item $m(m-1)(m-2)$ flags of type $F_{ABCC}$ with $A,B,C$ all distinct,
\be
F_1=(-u_{1}+\nu_A v_{3},-u_{1}+ \nu_B v_{2},-u_{1}+\nu_C v_{1},-u_{2}+\nu_A v_{3}) 
\ee
contributing  $- \frac{1}{12}$ as $y\to 1$.

\item $3 \times m(m-1)(m-2)$ flags of type $F_{ABCC}$ with $A,B,C$ all distinct,
\be
\begin{array}{ccccc}
F_2=(-u_{2}+\nu_A v_{3},-u_{1}+\nu_B v_{2},-u_{1}+\nu_C v_{1},-u_{1}+\nu_A v_{3})\\ 
F_3=(-u_{1}+\nu_C v_{3},-u_{2}+\nu_A v_{2},-u_{1}+\nu_B v_{1},-u_{1}+\nu_A v_{2})\\ 
F_4=(-u_{1}+\nu_C v_{3},-u_{1}+\nu_B v_{2},-u_{2}+\nu_A v_{1},-u_{1}+\nu_A v_{1})\\
 \end{array}
\ee
contributing $+ \frac{1}{12}$ as $y\to 1$; note that $F_1$ and
$F_2$ only differ by the order of the hyperplanes, while $F_{2,3,4}$ differ by a permutation
of $\{v_1,v_2,v_3\}$.

\item $6 \times m(m-1)^3$ flags of type $W_{ABCD}$ with $A\neq B,B\neq C,C\neq D$, 
\be
\begin{array}{ccccc}
W_1=(-u_{2}+\nu_A v_{3},-u_{1}+\nu_D v_{1},-u_{1}+\nu_C v_{2},-u_{2}+\nu_B v_{2})\\ 
W_2=(-u_{2}+\nu_A v_{3},-u_{1}+\nu_D v_{2},-u_{1}+\nu_C v_{1},-u_{2}+\nu_B v_{1})\\ 
W_3=(-u_{2}+\nu_A v_{2},-u_{1}+\nu_D v_{1},-u_{1}+\nu_C v_{3},-u_{2}+\nu_B v_{3})\\ 
W_4=(-u_{1}+\nu_D v_{3},-u_{2}+\nu_A v_{1},-u_{1}+\nu_C v_{2},-u_{2}+\nu_B v_{2})\\ 
W_5=(-u_{1}+\nu_D v_{2},-u_{2}+\nu_A v_{1},-u_{1}+\nu_C v_{3},-u_{2}+\nu_B v_{3})\\ 
W_6=(-u_{1}+\nu_D v_{3},-u_{2}+\nu_A v_{2},-u_{1}+\nu_C v_{1},-u_{2}+\nu_B v_{1})\\ 
 \end{array}
\ee
contributing $+ \frac{1}{12}$ as $y\to 1$. These
flags differ by permutations of $\{u_1,u_2\}$ and $\{v_1,v_2,v_3\}$.
\end{itemize}
In total, we thus find $2m(3m^3-7 m^2+3 m+1)$ stable flags, contributing as $y\to 1$
\be
\label{K23m}
m(m-1)(m-2)\times\frac{-1}{12} + 3 \times m(m-1)(m-2)\times\frac{1}{12} + 6 \times m(m-1)^3\times\frac{1}{12} = \frac{m^4}{2}-\frac{4 m^3}{3}+m^2-\frac{m}{6}
\ee
More generally, for $y\neq 1$ we find
 \be
\label{K23rei}
\chi_Q= \frac{y^{6m-4} +y^{4-6m}+\left(y+1/y\right)^4 
-\left(y^6+3   y^4+3 y^2+2\right) y^{2 m-4}
-\left(2 y^6+3 y^4+3 y^2+1\right) y^{-2-2m} }
   {\left(y-1/y\right)^4 \left(y+1/y\right)^2
   \left(y^2+1+y^{-2}\right)} 
\ee
in agreement with  Reineke's formula. 

\medskip

In the absence of flavor fugacities, the non-degenerate intersections collide into a single degenerate intersection involving all chiral multiplet hyperplanes. We find that 4 stable flags contribute, corresponding to $F_1,F_2,F_3,F_4$ above with $\nu_A=1$, with the flags $F_{2,3,4}$ producing equal contributions. As an example, for $m=3$ we find
\be
\label{K233}
\begin{split}
 &-\tfrac12\left( \frac{7}{y^6}+\frac{27}{y^4}+\frac{55}{y^2}+69+55y^2+27y^4+7y^6\right)
+\tfrac{3}{6} \left( \frac{9}{y^6}+\frac{29}{y^4}+\frac{61}{y^2}+75+61y^2+29y^4+9y^6\right)
\\
&=\frac{1}{y^6}+\frac{1}{y^4}+\frac{3}{y^2}+3+3y^2+y^4+y^6  
\qquad \stackrel{y\to 1}{\to} \qquad -\frac{247}{2}+3\times\frac{91}{2} = 13
\end{split}
\ee
again in agreement with \eqref{K23rei}.

\medskip

Let us now apply the Cauchy-Bose identity for the $U(2)$ vector multiplets, keeping the fugacities generic. In this case, the same non-degenerate intersections contribute separately to each of the two partitions $(1+1,2$):
\begin{itemize}
\item  Each of the $m(m-1)(m-2)$ flags of type $F_1$ contribute $(\frac14,-\frac13)$ 
\item Each of the $3 \times m(m-1)(m-2)$ flags of type $F_{2,3,4}$ contribute $(0,\frac{1}{12})$
\item Each of the  $6 \times m(m-1)^3$ flags of type $W_{1,\dots 6}$ contribute $(\frac{1}{12},0)$
\item In addition, there are $3\times2\times m(m-1)$ flags of type $F_{AABB}$ with $A\neq B$ which contribute $(\frac{1}{12},-\frac{1}{12})$ 
\end{itemize}
After summing these contributions, we arrive at the result predicted by the MPS formula \eqref{MPSpartial} with 
$S=\{1\}$,
\be
\label{K23mMPS}
\chi_{K_m(2,3)} = \frac{y-y^{-1}}{2(y^{2}-y^{-2})} 
\chi_{K_{2m}(1,3)}  + \frac{1}{2}
\chi_{S_{m,m}(3,1,1)}
\ee
where $K_{2m}$ is the Kronecker quiver with $2m$ arrows, while 
$S_{m,m}$ is the star quiver considered in \S\ref{sec_star} for $a_1=a_2=m$. Similarly,
applying the Cauchy-Bose identity for the $U(3)$ vector multiplets, one finds
\be
\chi_{K_m(2,3)} = \frac{y-y^{-1}}{3(y^{3}-y^{-3})} \chi_{K_{3m}(2,1)}  
+  \frac{y-y^{-1}}{2(y^{2}-y^{-2})}  \chi_{S_{2m,m}(2,1,1)}
+\frac{1}{6}   \chi_{S_{m,m,m}(2,1,1,1)}
\ee
in agreement with the MPS formula \eqref{MPSpartial} with $S=\{2\}$.

\medskip

Finally, let us apply the Cauchy-Bose identity for both the $U(2)$ and $U(3)$ vector multiplets. 
We find contributions from the 6 possible partitions $\lambda=(\lambda_1,\lambda_2)$ of the dimension vector $(2,3)$:
\begin{itemize}
\item For $\lambda=(2,3)$, we find $m$ flags of type $F_{AAAA}$ contributing $-\frac23$ each,
and $3m$ flags of type $F_{AAAA}$ contributing $\frac16$, leading to  $\chi_Q^\lambda = -m/6$ in the limit $y\to 1$;
\item For $\lambda=(2,2+1)$, we find $m^2$ flags of type $F_{AABB}$ contributing $+1$,
and $2m^2$  flags of type $F_{AABB}$ contributing $-\frac14$, leading to $\chi_Q^\lambda = m^2/2$;
\item For $\lambda=(2,1+1+1)$, we find $m^3$ flags of type $F_{ABCC}$ contributing $-\frac13$,
leading to $\chi_Q^\lambda = -m^3/2$;
\item For $\lambda=(1+1,3)$, we find $m^2$ flags of type $F_{AAAB}$ contributing $\frac12$,
leading to $\chi_Q^\lambda = m^2/2$;
\item For $\lambda=(1+1,2+1)$, we find 
$2m(m-1)$ flags of type $F_{AAAB}$ with $A\neq B$ 
and $m$ flags of type $F_{AAAA}$, each contributing $-\frac34$;  
$2m(m-1)$ flags of type $F_{AAAB}$ with $A\neq B$ 
and $m$ flags of type $F_{AAAA}$, each contributing $-\frac14$; and 
$2m(m-1)^2$ flags of type $W_{AABC}$ with $A\neq B$, $B\neq C$, each contributing $-\frac12$;
in total, $\chi_Q^\lambda = -m^3$;
\item For $\lambda=(1+1,1+1+1)$, we find $m^3$ flags of type $F_{ABCC}$ contributing $\frac14$;
$m(m-1)$ flags of type $F_{AAAB}$ contributing $\frac14$;  
$3m^3$  flags of type $F_{ABCC}$ plus
$3m(m-1)$ flags of type $F_{AAAB}$ with $A\neq B$, each contributing  $\frac1{12}$;
$6m(m-1)^2$ flags of type $W_{AABC}$
with $A\neq B, B\neq C$ contributing $\frac16$;  
$6m(m-1)^3$ flags of type $W_{ABCD}$ with $A\neq B, B\neq C,C\neq D$ contributing $\frac1{12}$;  in total, $\chi_Q^\lambda = m^4/2$;
\end{itemize}
These results are summarized in the table below, retaining the dependence on $y$:
\be
\begin{array}{|c|c|c|}
\hline
\lambda &  \chi_Q^\lambda & y\to 1 
\\
\hline
(2,3) 
& -\kappa(6m)\, \rho_2 \rho_3
& -m/6 
\\
 (2,2+1)
  & \kappa(2m) \kappa(4m) \rho_2^2
&m^2/2 
 \\
(2,1+1+1) 
&- \frac16 \kappa(2m)^3 \rho_2
& -m^3/3 
\\
 (1+1,3) 
& \frac12 \kappa(3m)^2 \rho_3 
& m^2/2 
\\
(1+1,2+1)
&- \frac12 \kappa(m) \kappa(2m) (\kappa(3m)+\kappa(m)) \rho_2
& -m^3 
\\
 (1+1,1+1+1)
  &\frac{1}{12} \kappa(m)^3 ( \kappa(3m) + 3\kappa(m)) 
  & m^4/2
  \\
  \hline
\end{array}
\ee
where $\rho_k=(y-1/y)/k/(y^k-y^{-k})$ and $\kappa(m)=(-1)^m(y^m-y^{-m})/(y-1/y)$. 
The contributions of the various partitions perfectly match 
the result of the Coulomb branch formula, which in this case follows from the wall-crossing formula
\cite[(A.4)]{Manschot:2010qz} (changing $m\to -m$ in this equation, and setting $\bOm(N_1,N_2)=0$
unless $N_1 N_2=0$)
\be
\label{Om23Coulomb}
\begin{split}
\chi_{K_m(2,3)} =&-\kappa(6m)\, \bOm(2,0)\, \bOm(0,3)
+ \frac12 [\kappa(3m)]^2\,  \bOm(1,0)^2\, \bOm(0,3) 
+\kappa(2m)\,\kappa(4m)  \bOm(2,0)\, \bOm(0,2)\, \bOm(0,1) \\
&- \frac12 \left[ \kappa(m)^2\,\kappa(2m) + \kappa(m) \kappa(2m)\, \kappa(3m) \right]\,  \bOm(1,0)^2\, \bOm(0,2)\, \bOm(0,1)\\
&- \frac16 \kappa(2m)^3\, \bOm(2,0)\, \bOm(0,1)^3
+\frac12 \left[ 3\kappa(m)^4 + \kappa(m)^3 \kappa(3m) \right]\,\bOm(1,0)^2\, \bOm(0,1)^3
\end{split}
\ee
where $\bOm(k,0)=\bOm(0,k)=\rho_k$. 

\subsubsection{Rank $(2,2)$}
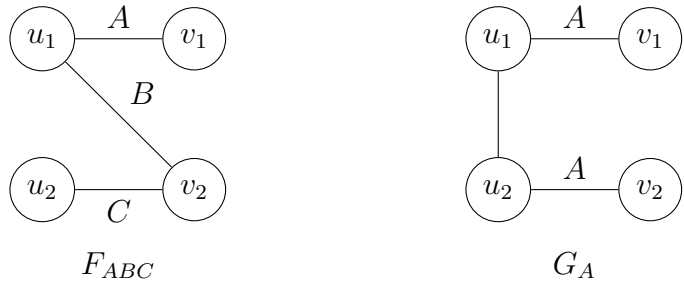
\begin{figure}[h]
\begin{center}
\begin{tikzpicture}[scale=1, minimum size=4mm]
\begin{scope}[shift={(-3,0)}]
  \node (u1) at ( -1,1) [circle,draw] {$u_1$};
  \node (u2) at ( -1,-1) [circle,draw] {$u_2$};
  \node (v1) at (1,1) [circle,draw] {$v_1$};
  \node (v2) at (1,-1) [circle,draw] {$v_2$};
  \node (c) at (0,-2) {$F_{ABC}$} ;
  \draw (u1) to node[auto] {$A$}  (v1);
   \draw (u1) to node[auto] {$B$}  (v2); 
   \draw (u2) tonode[auto,swap] {$C$}   (v2); 
\end{scope}
\begin{scope}[shift={(3,0)}]
  \node (u1) at ( -1,1) [circle,draw] {$u_1$};
  \node (u2) at ( -1,-1) [circle,draw] {$u_2$};
  \node (v1) at (1,1) [circle,draw] {$v_1$};
  \node (v2) at (1,-1) [circle,draw] {$v_2$};
  \node (c) at (0,-2) {$G_A$} ;
  \draw (u1) to node[auto] {$A$}  (v1);
   \draw (u1) to  (u2); 
   \draw (u2) tonode[auto] {$A$}   (v2); 
\end{scope}
\end{tikzpicture}
\end{center}
\caption{Graphs associated to the flags contributing to the equivariant $\chi$-genus of the 
Kronecker quiver $K_m(2,2)$ (up to permutations of $\{u_s\}$ and $\{v_{s'}\}$)
\label{fig:flags22}}
\end{figure}

In the case where the dimension vector is not primitive, there can be contributions from singularities involving vector multiplet hyperplanes. As a result, the contributions are still classified by trees but they are no longer bipartite, since they can involve edges of the form $e_s-e_{s'}$ or $f_s-f_{s'}$ corresponding to the hyperplanes $u_s/y-y\, u_{s'}$ or $v_s/y-y\, v_{s'}$. As an example, we consider the rank $(2,2)$ case. In the absence of flavor fugacities, we find that among the 21 singularities, 
three intersections support stable flags with non-trivial contributions. As an example, for $m=3$
we get
\be
\begin{array}{|c|c|}
\hline
\mbox{flag} & \mbox{residue} \\
\hline
F=( v_{2}-u_{1}, v_{1}-u_{1}, v_{2}-u_{2} )&  
-\frac12(\frac{2}{y^5}+\frac{11}{y^3}+\frac{20}{y}+20y+11y^3+2y^5)\\
G_1=( v_{1}-u_{1}, v_{2}-u_{2}, u_{2}-y^2 u_{1}  )& 
\frac{10+27 y^2+37 y^4+27 y^6+10 y^8}{4y^3(1+y^2)}  \\ 
G_2=(v_2-u_{1}, v_{1}-u_{2}, u_{2}-y^2 u_{1}  ) &  
\frac{10+27 y^2+37 y^4+27 y^6+10 y^8}{4y^3(1+y^2)}  \\
\hline
\end{array}
\ee
Adding up these terms, we find
\be
\bar\chi_{K_3(2,2)} =-\frac{(2+3y^2+2y^4)(1+y^4+y^8)}{2y^5(1+y^2)} 
\stackrel{y\to 1}{\rightarrow} -\frac{21}{4}
\ee
in agreement with Reineke's formula's for the rational invariant,
\be
\label{K22rei}
\bar\chi_{K_m(2,2)} =
-\tfrac{ \left(y^{m}-y^{-m}\right) \left(2 y^{3 m-2}+2 y^{2-3m}-2 y^{m+2}-2 y^{-2-m} -y^{3}-y^{-3}
+y+y^{-1}\right)}{2
   \left(y-1/y\right)^3 \left(y+1/y\right)^2}
\stackrel{y\to 1}{\rightarrow} \frac{-m^3}{2}+m^2-\frac{m}{4}
\ee
For generic flavor fugacities, the singularities become non-degenerate.  We find that
the following stable flags contribute:
(see Figure \ref{fig:flags22})
\begin{itemize}
\item $2\times m(m-1)^2$ flags of type $F_{ABC}$ with $A\neq B, B\neq C$,
\be
\begin{split}
F_1=& ( -u_{1}+ \nu_B\, v_{2}, -u_{1}+\nu_A\, v_1,-u_{2} +\nu_C  v_{2})  
\\
F_2 =&  ( -u_{1}+ \nu_A\, v_{1}, -u_{2}+\nu_C\, v_2,-u_{1} +\nu_B  v_{2}) 
\end{split}
\ee
 contributing $-\frac14$ as $y\to 1$;
\item $2m$ flags of type $G_A$,
\be
\begin{split}
G_1=&  ( -u_{1}+ \nu_A\, v_{1}, -u_{2}+\nu_A\, v_2,-u_{1} y^2+ u_{2}) \\
G_2 =& ( -u_{1}+ \nu_A\, v_{2}, -u_{2}+\nu_A\, v_1,-u_{1} y^2  +u_{2}) 
\end{split}
\ee
 contributing $\frac18$ in the limit $y\to 1$ as $y\to 1$;
\end{itemize}
In total, we get 
 \be
 \bar\chi_{K_m(2,2)} = 2m(m-1)^2 \times -\frac14 + 2m\times \frac18 = -\frac{m^3}{2} + m^2 - \frac{m}{4}
 \ee
 in agreement with \cite[\S 6.3]{Ohta:2014ria}.
 
Applying the Cauchy-Bose formula for both nodes, this result decomposes as a sum over partitions
$\lambda=(\lambda_1,\lambda_2)$
of the dimension vector $(2,2)$:
\begin{itemize}
\item For $\lambda=(2,2)$, we find $m$ flags of type $F_{AAA}$ contributing $-\frac12$ each
and $2m$ flags of type $G_{A}$ contributing $-\frac18$ each, leading to $\chi_Q^\lambda 
= -\frac{m}{4}$;
\item For $\lambda=(2,1+1)$ or $\lambda=(1+1,2)$, we find $m^2$ flags of type $F_{AAB}$ contributing $\frac12$ each, leading to $\chi_Q^\lambda = \frac{m^2}{2}$;
\item For $\lambda=(1+1,1+1)$, we find $2m(m-1)$ flags of type $F_{AAB}$ with $A\neq B$ and 
$m$ flags of type $F_{AAA}$, contributing $-\frac12$ each; and $2m(m-1)^2$ flags of type
$F_{ABC}$ with $A\neq B\neq C$; in total $\chi_Q^\lambda = -\frac{m^3}{2}$;
\end{itemize}
These results are summarized in the table below, retaining the dependence on $y$:

\be
\begin{array}{|c|c|c|}
\hline
\lambda &  \chi_Q^\lambda & y\to 1 
\\
\hline
(2,2) &\kappa(4m) \,\rho_2^2 & -m/4 \\
 (2,1+1)&  \frac12 [\kappa(2m)]^2\, \rho_2 & m^2/2 \\
(1+1,2) & \frac12 [\kappa(2m)]^2\, \rho_2   & m^2/2  \\
(1+1,1+1) &  \frac14 \kappa(m)^2\kappa(2m) & -m^3/2 \\
\hline
\end{array}
\ee
The contributions of the various partitions match 
the result of the Coulomb branch formula \cite[(A.4)]{Manschot:2010qz}
\be
\begin{split}
 \bar\chi_{K_m(2,2)}=&-\kappa(4m)\, \bOm(2,0)\, \bOm(0,2)
+ \frac12 [\kappa(2m)]^2\,  \bOm(1,0)^2\, \bOm(0,2) 
+ \frac12 [\kappa(2m)]^2\,  \bOm(2,0)\, \bOm(0,1)^2\\&
-\frac14 \kappa(m)^2\kappa(2m)\,  \bOm(1,0)^2\, \bOm(0,1)^2
\end{split}
\ee
where $\bOm(k,0)=\bOm(0,k)=\rho_k$. 

\subsection{Non-abelian quivers with oriented loops}

Finally, we turn to an example of non-abelian quiver with oriented loops. In this case, applying the Cauchy-Bose formula to all nodes leads to a sum over partitions of the total dimension vector 
as a sum of multiples of basis vectors $\ell \alpha_a$, whereas the Coulomb branch formula also 
includes contributions of single-centered invariants involving combinations of these basis vectors which support an oriented loop. Upon matching the two formulae, we find that the single-centered invariants naturally decompose into contributions from different partitions, whose mathematical meaning remains to elucidate.

To exhibit this phenomenon, let us consider the simple case of a 3-node cyclic quiver $C_{a,b,c}$ with 
$a>0, b>0, c>0$ and dimension vector $(2,1,1)$. 
In order to compare with the analysis in Sec 6.1 of \cite{Manschot:2012rx} (up to a cyclic permutation of the nodes), we assume
\be
\begin{split}
&b<2a, \quad c<a, \quad k=b+2(c-a)>0, \\
&\zeta_2>0, \quad \zeta_2+\zeta_3>0, \quad \zeta_3<0, \quad \zeta_1\to 0^-
\end{split}
\ee
which includes   the attractor point  $\zeta^*=(c-a,2a-b,b-2c)$. 
Simple choices of $(a,b,c)$ satisfying these conditions are
\be
(4,7,3), (5, 7, 4), (5, 8, 4), (6, 7, 5), (6, 8, 5), (7, 7, 6),  (7,8,6), (8, 7, 7), (8, 8, 7), ...
  \ee
We upgrade the stability vector $\zeta$ into  $(\eta_0,\eta_1;\eta_2,\eta_3) = (\zeta_1 -\epsilon, \zeta_1+\epsilon, \zeta_2, \zeta_3)$ with
$0<\epsilon\ll |\zeta_1| $, and choose the R-charges to be $R_{12}=R_{23}=R_{31}=2/3$ 
so as to allow for a generic superpotential. 
Denoting the Cartan variables by $(v_1,v_2;u_1,u_2)$,
the $\chi$-genus of the quiver moduli space is given by a sum of residues of 
\be
\begin{split}
Z_Q =& \frac{1}{2(y-1/y)^3} 
\frac{ \left(v_1-v_2\right) \left(v_2-v_1\right)}{\left(v_2/y-v_1 y\right) \left(v_1/y-v_2 y\right)}
\times \left( \frac{y^{1/3} u_2- u_1/y }{u_1-y^{-2/3} u_2}\right)^b
\\
&\times 
\prod_{s=1,2} 
\left( \frac{y^{1/3} u_1- v_s/y }{v_s-y^{-2/3} u_1}\right)^a  
\left( \frac{y^{1/3} v_s- u_2/y }{u_2-y^{-2/3} v_s}\right)^c
\frac{\de u_1}{u_1} \frac{\de u_2}{u_2} \frac{\de v_1}{v_1} 
   \end{split}
\ee 
Out of the 15 singular points, 12 intersections are non-degenerate and three are degenerate. Out of those three, one is projective and two are not, but have $\det\kappa=0$ so can be safely ignored. For the stability conditions above, the only contributing stable flag is $(u_1-y^{-2/3} u_2,u_2-y^{-2/3} v_1,u_2-y^{-2/3}v_2)$, producing a symmetric Laurent polynomial of degree $k-5$, in agreement with the complex dimension of $\cM_Q$. On the other hand, the Coulomb branch
formula in this chamber predicts
\be
\label{Coulomb211}
\begin{split}
\chi_Q=& \OmS(2,1,1) 
+ \left( \frac{y-1/y}{2(y^2-1/y^2)} g_C(\{2\gamma_1,\gamma_2,\gamma_3\}) 
+ H( \{\gamma_1,\gamma_2,\gamma_3\},\{2,1,1\})  \right) 
\\&
+ \left( \frac12 g_C(\{\gamma_1,\gamma_1,\gamma_2,\gamma_3\}) + 
H( \{\gamma_1,\gamma_1,\gamma_2,\gamma_3\},\{1,1,1,1\})  \right)
\end{split}
\ee
where $\OmS(2,1,1)$ is the single-centered invariant, and the modified Coulomb indices are 
given by  \cite[\S6.1]{Manschot:2012rx} 
\be
\begin{split}
g_C(\{2\gamma_1,\gamma_2,\gamma_3\})=&
\frac{(-1)^k (y^k+y^{-k})}{(y-1/y)^2} 
\\
H( \{\gamma_1,\gamma_2,\gamma_3\},\{1,1,2\}) 
=& \begin{cases}{1\over 4} (y - y^{-1})^{-2} (y+y^{-1})^{-1}
\left\{-(y+y^{-1})^2 + (-1)^{k/2} (y-y^{-1})^2 \right\}
\cr
{1\over 2}  \, (y-y^{-1})^{-2} 
\end{cases} 
\\
 g_C(\{\gamma_1,\gamma_1,\gamma_2,\gamma_3\})
=&\frac{(-1)^{k+1}(y^k-y^{-k})}{(y-1/y)^3}
\\
H( \{\gamma_1,\gamma_2,\gamma_3,\gamma_3\},\{1,1,1,1\})
=& \begin{cases} {1\over 4}\, \, k\, (y - y^{-1})^{-2} (y+y^{-1})
\\
-{1\over 2}\, \, k\, (y - y^{-1})^{-2} 
\end{cases}
\end{split}
\ee
where the two options correspond to $k=b+2(c-a)$ even and odd. respectively.
By comparing the result from the residue formula with \eqref{Coulomb211}, we can read off
the single-centered invariant  $\OmS(2,1,1)$, e.g. for $(a,b,c)=(7,8,6)$ we get 
\be
\Omega(2,1,1)=1862(y+1/y) \quad \rightarrow \quad \OmS(2,1,1)=1863(y+1/y)
\ee

\begin{figure}[h]
\begin{center}
\begin{tikzpicture}[scale=2, minimum size=4mm]
\begin{scope}[shift={(-2,0)}]
   \node (a) at ( 0,.7) [circle,draw] {$1$};
  \node (b) at ( 1,-1) [circle,draw] {$1$};
  \node (c)  at ( -1,-1) [circle,draw] {$1$};
 \draw [->] (a) to node[auto] {$2a$} (b);
 \draw [->] (b) to node[auto,swap] {$b$} (c);
 \draw [->] (c) to node[auto] {$2c$} (a);
 \end{scope}
 \begin{scope}[shift={(+2,0)}]
   \node (a) at ( -.5,.7) [circle,draw] {$1$};
  \node (b) at (.5,.7) [circle,draw] {$1$};
  \node (c)  at ( 1,-1) [circle,draw] {$1$};
   \node (d)  at (-1,-1) [circle,draw] {$1$};
 \draw [->] (a) to node[auto] {$a$} (c);
  \draw [->] (b) to node[auto] {$a$} (c);
 \draw [->] (d) to node[auto,swap] {$c$} (a);
 \draw [->] (d) to node[auto,swap] {$c$} (b);
  \draw [->] (c) to node[auto,swap] {$b$} (d);
  \end{scope}
\end{tikzpicture}
\end{center}
\caption{Abelian quivers rising in the MPS formula for the 3-node quiver with rank $(2,1,1)$.
\label{fig:MPS211}}
\end{figure}
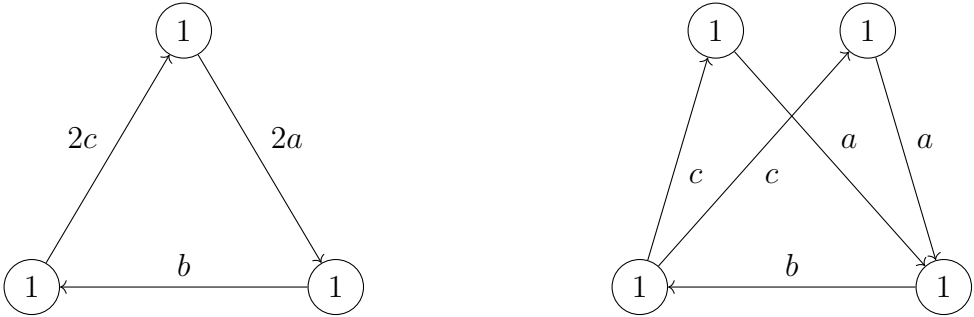

Now, applying the Cauchy-Bose formula for the $U(2)$ vector multiplets, the $\chi$-index naturally splits into two terms $\chi_Q^{(2)}$, $\chi_Q^{(1+1)}$ and associated to the partitions of $N=2$. 
From the Coulomb branch picture, these two partitions correspond to bound states of charges 
$\{2\gamma_1,\gamma_2,\gamma_3\}$
and $\{\gamma_1,\gamma_1,\gamma_2,\gamma_3\}$, corresponding to the two terms in bracket
in \eqref{Coulomb211}. Thus, it is natural to define 
`partial' single-centered invariants $\OmS^{(2)}(2,1,1)$ and $\OmS^{(1+1)}(2,1,1)$ via 
\be
\begin{split}
\chi_Q^{(2)} =& 
\OmS^{(2)}(2,1,1) + \frac{y-1/y}{2(y^2-1/y^2)} g_C(\{2\gamma_1,\gamma_2,\gamma_3\}) 
+ H( \{\gamma_1,\gamma_2,\gamma_3\},\{2,1,1\})  
\\
\chi_Q^{(1+1)} =& 
\OmS^{(1+1)}(2,1,1) +  \frac12 g_C(\{\gamma_1,\gamma_1,\gamma_2,\gamma_3\}) + 
H( \{\gamma_1,\gamma_1,\gamma_2,\gamma_3\},\{1,1,1,1\}) 
\end{split}
\ee
such that $\OmS(2,1,1)=\OmS^{(2)}(2,1,1)+\OmS^{(1,1)}(2,1,1)$. E.g. for $(a,b,c)=(7,8,6)$,
\be
\begin{split}
\OmS^{(2)}(2,1,1) =&- 3 \left(y+1/y\right) \left(1903 y^2+10976 +1903 y^{-2} \right) \\
\OmS^{(1+1)}(2,1,1) =&3
   \left(y+1/y\right) \left(1903 y^2+11597+1903 y^{-2}\right)
   \end{split}
\ee
which correctly add up to $1863(y+1/y)$.
One might expect these
two contributions to be related to the single-centered invariants of the 3-node and 4-node
Abelian quivers obtained by applying the MPS formula \eqref{MPSpartial} to the node $V_1$ 
(see Figure \ref{fig:MPS211}).
This expectation is indeed borne out for $\OmS^{(1,1)}(2,1,1)$, which is equal to half of the single-centered invariant $\OmS(1,1,1,1)$ of the 4-node Abelian quiver, but not for $\OmS^{(2)}(2,1,1)$ which appears to be unrelated to the single-centered invariant $\OmS(1,1,1)$ associated to the 3-node Abelian quiver $C_{2a,b,2c}$. It would be interesting to understand the mathematical significance of $\OmS^{(2)}(2,1,1)$, or similar `partial' single-centered invariants arising at higher rank. 
More generally,  it would be of great interest to find a residue prescription (or otherwise) for
computing the single-centered invariants $\OmS(\gamma,y)$ directly.

\medskip

\acknowledgments

The authors are grateful to M. Reineke and T. Weist for helpful correspondence.
S. M. acknowledges the hospitality of LPTHE during part of this project, and
the J C Bose Fellowship of Rajesh Gopakumar, from the SERB, Govt. of India.
The research of B.P.  is supported in part by French state funds
managed by the Agence Nationale de la Recherche (ANR) in the context of the LABEX ILP (ANR-11-IDEX-0004-02, ANR-10- LABX-63).

\appendix

\section{Implementation of the residue formula in {\tt CoulombHiggs.m} \label{sec_math}}

The computations reported in this paper were in part carried out using the Mathematica package 
{\tt CoulombHiggs.m} by the last-named author, which was first released along with \cite{Manschot:2013sya} and has since then been extended to include an implementation of the residue 
formula \eqref{Omloc} for general non-Abelian quivers. This package can be freely downloaded
from the last author's home page\footnote{{\tt http://www.lpthe.jussieu.fr/$\sim$\,pioline/computing.html}}.
Below we briefly outline how to use this package to reproduce some of the computations in this paper, referring to the documentation for more complete information. 

Assuming that the file {\tt CoulombHiggs.m} is present in the user's {\sc Mathematica} Application 
directory, the package is loaded by entering 

\mathematica{1}{1.0}{ <<CoulombHiggs`}{CoulombHiggs 5.0 - A package for evaluating quiver invariants. }

To specify to the Kronecker quiver $K_3(2,3)$ considered in \S\ref{sec_k23}, we first feed in the adjacency matrix, R-charge matrix, stability parameter, dimension vector to the routine \fun{JKInitialize}:

\mathematica{2}{1.0}{Mat=\{\{0, -3\},\{3,0\}; RMat=0*Mat; Cvec=\{-1/2,1/3\};Nvec=\{2,3\}; 
JKInitialize[Mat, RMat, Cvec, Nvec]; 
JKChargeMatrix // MatrixForm}
{$\left(
\begin{array}{ccccccc}
 -1 & 0 & 1 & 0 & 0 & 0 & 3 \\
 0 & -1 & 1 & 0 & 0 & 0 & 3 \\
 -1 & 0 & 0 & 1 & 0 & 0 & 3 \\
 0 & -1 & 0 & 1 & 0 & 0 & 3 \\
 -1 & 0 & 0 & 0 & 1 & 0 & 3 \\
 0 & -1 & 0 & 0 & 1 & 0 & 3 \\
\end{array}
\right)$}

Among other things, this routine constructs the matrix \var{JKChargeMatrix}, whose rows contain the 
charges of the chiral multiplets under the Cartan algebra $U(1)^{2}\times U(1)^3$, as well as their 
R-charge (here, 0) and multiplicity (here, 3); as well as the vector \var{JKEta}, which is a small
perturbation of $(\zeta_1,\zeta_1,\zeta_2,\zeta_2,\zeta_2)$ with increasing entries for each node. 
The integrand
$Z_Q$ in \eqref{Zuhbar} can be displayed by calling \fun{ZTrig[JKChargeMatrix, Nvec]}, or 
\fun{ZEuler[]}, \fun{ZRational[]}, \fun{ZElliptic[]} for \eqref{Zuhbar1},  \eqref{Zv} or \eqref{Zelliptic}, respectively. For computing the index of non-quiver type systems, the charge matrix can be specified by hand.

Having constructed the charge matrix and generalized stability vector, we can then feed these data into the routine $\fun{JKIndex}$,

\mathematica{3}{1.0}{
JKIndex[JKChargeMatrix, Nvec,JKEta];
JKChiGenus 

{\sl 66 stable flags in total; from computing the Euler number, 4 stable flags appear to contribute}
}{$\left\{-\frac{7 y^{12}+27 y^{10}+55 y^8+69 y^6+55 y^4+27 y^2+7}{2 y^6},
  -\frac{-9 y^{12}-29 y^{10}-61 y^8-75 y^6-61 y^4-29 y^2-9}{6 y^6} \right., $
  $\left. -\frac{-9 y^{12}-29 y^{10}-61 y^8-75 y^6-61 y^4-29 y^2-9}{6 y^6},
   -\frac{-9 y^{12}-29 y^{10}-61 y^8-75 y^6-61 y^4-29 y^2-9}{6 y^6} \right\}$}
   
 Each of the entries in the result, stored in the global variable  \var{JKChiGenus}, gives the 
 contribution to the $\chi$-genus of those stable flags which contribute non-trivially to the Euler
 number. In this case, out of 66 stable flags,  4 of those give non-trivial contributions, reproducing the result in \eqref{K233}. The relevant flags are stored in the global variable \var{JKRelevantStableFlags},

\mathematica{4}{1.0}{JKListuDisplay = \{u1, u2, v1, v2, v3\};
DisplayFlagList[JKRelevantStableFlags]
}
{$\left(
\begin{array}{ccc}
 \{0,0,0,0\} &
   \{\text{v3}-\text{u1},\text{v2}-\text{u1},\text{v1}-\text{u1},\text{v3}-\text{u2}\} & -1 \\
 \{0,0,0,0\} &
   \{\text{v3}-\text{u2},\text{v2}-\text{u1},\text{v1}-\text{u1},\text{v3}-\text{u1}\} & -1 \\
 \{0,0,0,0\} &
   \{\text{v3}-\text{u1},\text{v2}-\text{u2},\text{v1}-\text{u1},\text{v2}-\text{u1}\} & -1 \\
 \{0,0,0,0\} &
   \{\text{v3}-\text{u1},\text{v2}-\text{u1},\text{v1}-\text{u2},\text{v1}-\text{u1}\} & -1 \\
\end{array}
\right)$
}
where the first entry in each row corresponds to the intersection point (dropping the coordinate along the frozen coordinate $u_1$, specified by the variable \var{JKFrozenCartan} defined by \fun{JKInitialize}), the second entry to the ordered list of hyperplanes defining each flag $F$, and the last entry to $\sign(\det\kappa_F)$. The covering tree associated to each flag can also be displayed by using
\fun{DisplayFlagTree},

\mathematica{5}{1.0}{JKVertexLabels = \{1 -> u1, 2 -> u2, 3 -> v1, 4 -> v2, 
  5 -> v3\}; Table[
 DisplayFlagTree[JKRelevantStableFlags[[i]]], {i, 
  Length[JKRelevantStableFlags]}]
}
{\includegraphics[width=10cm,bb=0 0 500 250]{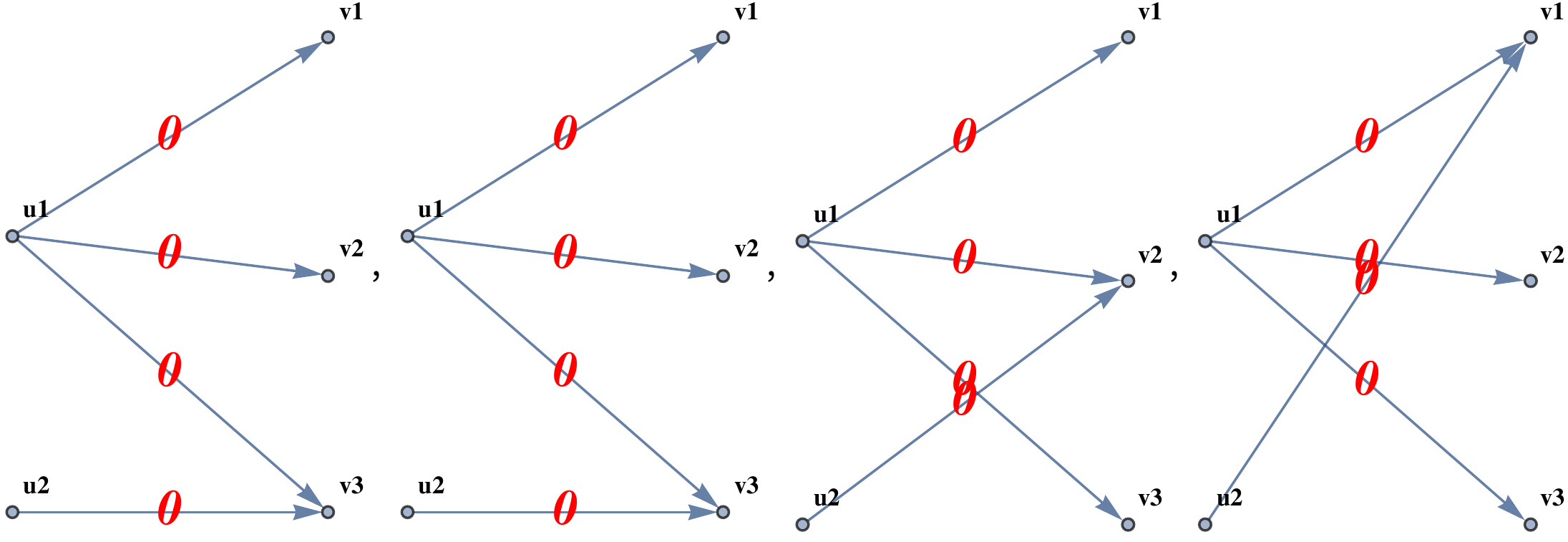}
}
  
The red zeros on each arrow indicate half the R-charge, which vanishes in this example. Flavor fugacities can be switched on by defining 

\mathematica{6}{1.0}{RMat = 2 FlavoredRMatrix[Mat]; JKInitialize[Mat,RMat, Cvec, Nvec]; }{}

The global variable \var{JKChargeMatrix} now includes $2\times 3\times 3$ rows, 
each of which with R-charge 
$2 {\rm th}[A]$ where ${\rm th}[A]$ are generic fugacity parameters. Running \fun{JKIndex} will now produce $1410$ stable flags, out of which $168$ contribute a non-zero residue as in \eqref{K23m}.

Finally, the same computation can be carried out using the Cauchy-Bose formula for either, or both factors of the gauge group $U(2)\times U(3)$. This is done using \fun{JKIndexSplit}, which takes 
the same arguments as \fun{JKIndex} plus the subset  $S\subset \{1,2\}$ as in \eqref{MPSpartial}.
E.g. the results reported above \eqref{K23mMPS} can be verified by using

\mathematica{7}{1.0}{
JKIndexSplit[JKChargeMatrix, Nvec,JKEta,\{1\}];
JKChiGenus 
}{}
and studying the structure of the relevant flags in \var{JKRelevantStableFlags}.

\providecommand{\href}[2]{#2}\begingroup\raggedright\endgroup


\end{document}